\documentclass[%
preprint, 
 amsmath,amssymb,
 aps, physrev,
reprint
]{revtex4-2}

\usepackage{graphicx}
\usepackage{dcolumn}
\usepackage{bm}
\usepackage{multirow}
\usepackage{amsmath,mathtools}
\usepackage{physics}
\usepackage{color}
\usepackage{hyperref}

\DeclareUnicodeCharacter{0301}{\'{e}}

\def\tivo{\text{Ti}_{\text{Zn}}\text{v}_\text{O}}
\def\nbvo{(\text{Nb}_{\text{Zn}}\text{v}_\text{O})^+}
\def\vvo{(\text{V}_{\text{Zn}}\text{v}_\text{O})^+}
\def\movo{(\text{Mo}_{\text{Zn}}\text{v}_\text{O})^{2+}}
\def\xvo{\text{X}_{\text{Zn}}\text{v}_\text{O}}

\def\vo{\text{v}_\text{O}} 
\def\mozn{(\text{Mo}_\text{Zn})^{2+}}
\def\gs{\prescript{3}{}{A}_2} 
\def\tes{\prescript{3}{}{E}} 
\def\sgs{\prescript{1}{}{E}} 
\def\fses{\prescript{1}{}{A}_1} 

\def\ra{\rightarrow} 
\def\aone{\text{a}_1}

\def\ttwo{\text{T}_2}

\begin{document}

\preprint{APS/123-QED}

\title{Deep Spin Defects in Zinc Oxide for High-Fidelity Single-Shot Readout}

\author{Shimin Zhang \footnotemark[1]}
\affiliation{Department of Materials Science and Engineering, University of Wisconsin-Madison, Madison, Wisconsin, 53706, USA}
\author{Taejoon Park \footnotemark[1]}
\thanks{SZ, TP, and EP contributed equally.}
\affiliation{SKKU Advanced Institute of Nanotechnology, Sungkyunkwan University, Suwon, Gyeonggi 16419, Korea}
\affiliation{Department of Energy Systems Research and Department of Physics,  Ajou University, Suwon, Gyeonggi 16499, Korea}
\author{Erik Perez \footnotemark[1]}
\thanks{SZ, TP, and EP contributed equally.}
\affiliation{Department of Materials Science and Engineering, University of Wisconsin-Madison, Madison, Wisconsin, 53706, USA}
\author{Kejun Li}
\affiliation{Department of Materials Science and Engineering, University of Wisconsin-Madison, 53706, USA}
\affiliation{Department of Physics, University of California, Santa Cruz, California, 95064, USA}
\author{Xingyi Wang}
\affiliation{Department of Electrical and Computer Engineering, University of Washington, Seattle, WA, 98195, USA}
\author{Masoud Mansouri}
\affiliation{Departamento de Qu\'imica, M\'odulo 13, Universidad Aut\'onoma de Madrid, Madrid, 28049, Spain}
\author{Yanyong Wang} 
\affiliation{School of Science \& Engineering, Tulane University, 6823 St Charles Ave, New Orleans, LA 70118, USA}
\author{Jorge D Vega Bazantes}
\affiliation{School of Science \& Engineering, Tulane University, 6823 St Charles Ave, New Orleans, LA 70118, USA}
\author{Ruiqi Zhang}
\affiliation{School of Science \& Engineering, Tulane University, 6823 St Charles Ave, New Orleans, LA 70118, USA}
\author{Jianwei Sun}
\affiliation{School of Science \& Engineering, Tulane University, 6823 St Charles Ave, New Orleans, LA 70118, USA}
\author{Kai-Mei C. Fu}
\email{kaimeifu@uw.edu}
\affiliation{Department of Physics, University of Washington, Seattle, WA, 98195, USA}
\affiliation{Department of Electrical and Computer Engineering, University of Washington, Seattle, WA, 98195, USA}
\affiliation{Physical Sciences Division, Pacific Northwest National Laboratory, Richland, Washington 99352, USA}
\author{Hosung Seo}
\email{seo.hosung@skku.edu}
\affiliation{SKKU Advanced Institute of Nanotechnology, Sungkyunkwan University, Suwon, Gyeonggi 16419, Korea}
\affiliation{Department of Quantum Information Engineering, Sungkyunkwan University, Suwon, Gyeonggi 16419, Korea}
\affiliation{Department of Energy Systems Research and Department of Physics,  Ajou University, Suwon, Gyeonggi 16499, Korea}
\affiliation{Center for Quantum Information, Korea Institute of Science and Technology, Seoul 02792, Korea}
\author{Yuan Ping}
\email{yping3@wisc.edu}
\affiliation{Department of Materials Science and Engineering, University of Wisconsin-Madison, Madison, Wisconsin, 53706, USA}
\affiliation{Department of Physics, University of Wisconsin-Madison, Madison, Wisconsin, 53706, USA}
\affiliation{Department of Chemistry, University of Wisconsin-Madison, Madison, Wisconsin, 53706, USA}

\date{\today}

\begin{abstract}
Wide-band gap oxides such as ZnO are favorable hosts for spin defect qubits due to their dilute nuclear spin background and potential for ultra-high purity. Yet, a deep-level defect qubit with robust optical and spin properties has not been identified in this material. Here, using first-principles calculations, we predict that the molybdenum–vacancy complex, $\movo$, exhibits the essential characteristics of an optically addressable spin qubit: a spin-triplet ground state, visible-range optical transitions with high quantum yield, and an unusually small Huang–Rhys factor ($\sim$5, compared to 10–30 in known ZnO defects). We further find long spin coherence times (T$_2$ $\sim$4 ms) when both nuclear and impurity spin baths are considered, with paramagnetic impurities setting a threshold concentration of 0.035 ppm. Importantly, the combination of strong spin–orbit coupling and the absence of Jahn–Teller distortion supports spin-selective intersystem crossing and high-fidelity single-shot readout at elevated temperatures and across wide magnetic field ranges. By identifying ZnO as a host for deep-level defect qubits, our work points toward a pathway to scalable, integrable oxide-based quantum technologies and broadens the material foundation for solid-state quantum information science.
\end{abstract}

\maketitle


\section{\label{sec_intro}Introduction}
Point defects in wide-band gap solids
with unpaired spins are emerging as promising candidates for quantum bits (qubits) due to their exceptional advantages such as elevated temperature operation and long coherence times. 
Consequently, the search for further spin qubit candidates has become a rapidly growing field, aiming to identify systems that can enhance existing quantum technologies or expand their capabilities.
These efforts have led to the discovery of group-IV-vacancy qubits in diamond~\cite{dohertyNitrogenvacancyColourCentre2013,orphal-kobinCoherentMicrowaveOptical,harrisCoherenceGroupIVColor2024,karapatzakisMicrowaveControlTinVacancy2024}, vacancy spins in SiC~\cite{tarasenkoSpinOpticalProperties2018,castellettoSiliconCarbideColor2020}, and more recently, boron vacancy spins in hexagonal boron nitride~\cite{gottschollInitializationReadoutIntrinsic2020a,gottschollRoomTemperatureCoherent2021a,gongCoherentDynamicsStrongly2023}. Notably, first-principles theoretical approaches have played a crucial role in this process by efficiently exploring the vast configuration space of defects in a material, narrowing down to the best candidates, and navigating the unknown connection between desirable qubit properties and the defect structure.

Finding optimal quantum defects for spin qubits requires evaluating both defect and host materials. An ideal host should have a wide band gap which can span the energy range of ground and excited defect levels, minimal nuclear spin density for long spin coherence times, and can be grown as high-quality single crystals for minimum unwanted defects and impurities~\cite{kanaiGeneralizedScalingSpin2022a}. Given a host material, the best qubit candidates are identified based on deep defect levels, high spin states, strong radiative recombination rates compared to non-radiative ones, large Debye-Waller factor, and long spin relaxation and coherence times~\cite{pingComputationalDesignQuantum2021,awschalomQuantumTechnologiesOptically2018,weberQuantumComputingDefects2010}.

ZnO has recently gained recognition as an excellent host material due to its exceptional purity, achievable through molecular beam epitaxy (MBE) with sub-ppb background impurity concentrations~\cite{falsonMagnesiumDopingControlled2011,liElectronMobilityZnMgO2013,falsonMgZnOZnOHeterostructures2016}. This minimizes unintentional defects and paramagnetic noise, crucial for long coherence times. Furthermore, ZnO benefits from the inherent properties of oxides, where oxygen is 99.7\% nuclear spin-free~~\cite{haynes2010crc}, ensuring a magnetically quiet environment  and further reducing noise. The piezoelectricity property of ZnO introduces the possibility of strain tuning of the quantum states of spin defects. 

Notably, several shallow donors in ZnO have already been identified as potential qubit candidates due to their advantageous optical and spin properties. Neutral indium (In) and gallium (Ga) donor-bound electrons ($D^{0}$) form spin-1/2 qubit systems, exhibiting long spin relaxation times ($T_1$) ($>$100ms)~\cite{niaourisEnsembleSpinRelaxation2021,linpengCoherencePropertiesShallow2018}, Hahn-spin-echo decoherence times ($T_2$ =$\sim$50$\mu s$)\cite{linpengCoherencePropertiesShallow2018}, and narrow inhomogeneous linewidth~\cite{wangPropertiesDonorQubits2023,viitaniemiCoherentSpinPreparation2022}. Experimentally, all-optical methods have been utilized to control donor spin qubit states at cryogenic temperatures~\cite{linpengCoherencePropertiesShallow2018}. 

Expanding ZnO spin qubits beyond shallow donors is crucial for advancing ZnO-based spin-photon interfaces and exploring new capabilities. The low binding energy of shallow donors requires low-temperature operation~\cite{linpengCoherencePropertiesShallow2018, wangPropertiesDonorQubits2023,NiaourisLinewidth2024}.
Additionally, their UV-range optical emission is unsuitable for long-distance communications by optical fibers. Therefore, identifying deep-level defect qubits in ZnO is essential for room-temperature optical initialization and quantum emission in the IR and visible range that can be coherently coupled to the spin. Unlike spin-1/2 shallow donors, deep-level defects with a highly localized spin-triplet ground state would offer greater robustness against environmental noise, making them ideal for quantum sensing. Interestingly, several previous studies reported strong photon emission from deep-level defects in ZnO~\cite{grayLocalAtomicEnvironment2017,haoStructuralOpticalMagnetic2012,liaoDcThermalPlasma2006,shahroosvandSolutionbasedSyntheticStrategies2013,singhInvestigationLowtemperatureExcitonic2010}, but no spin qubits have been identified yet.

High-fidelity readout of spins is vital for spin-based quantum applications. 
Single-shot readout enables 
high-fidelity and efficient measurement, essential for quantum communication and algorithm execution~\cite{irberRobustAllopticalSingleshot2021,christleIsolatedSpinQubits2017,hopperSpinReadoutTechniques2018}. A readout fidelity above 79\%, corresponding to a signal-to-noise ratio greater than one, is widely accepted as the threshold for reliable single-shot detection~\cite{hopperSpinReadoutTechniques2018}. This technique relies on resonant excitation of spin-selective, spin-conserving optical cycling transitions that are well isolated from intersystem crossings (ISC), thereby enhancing both contrast and photon yield~\cite{robledoHighfidelityProjectiveReadout2011}. However, fidelity can be limited by spin leakage, where unwanted spin-flip transitions occur in the excited state due to magnetic noise or vibronic interactions. Previous studies have shown that reduced Jahn--Teller distortion, weak spin--phonon interactions, and large spin sublevel splittings help suppress spin leakage in systems such as SnV, SiV center in diamond, and divacancies in 4H--SiC~\cite{rosenthalSingleShotReadoutWeak2024,christleIsolatedSpinQubits2017,andersonFivesecondCoherenceSingle2022,Koehl2011-iv,Sukachev2017-rr}.

In this article, we investigate nitrogen vacancy (NV) center like complex vacancy defects in ZnO, and identify the molybdenum vacancy complex ($\movo$) as a promising spin qubit candidate with superior optical and spin properties. It features a highly-polarized intersystem crossing with large non-axial SOC components and zero axial ones. Its strong spin–orbit–induced spin splitting in the excited state, along with a suppressed Jahn–Teller effect, allows minimal spin mixing even under magnetic fluctuations and elevated temperatures. Optically, it exhibits a remarkably low Huang–Rhys factor ($\sim5$), significantly lower than other known ZnO defects (typically $>10$), resulting in a sharp zero-phonon line and high quantum yield due to suppressed electron–phonon coupling and phonon-assisted nonradiative processes. In addition, a long spin decoherence time($T_2$) in ms order has been found, which could be limited by paramagnetic impurities even after isotopic purifications. These properties result in a promising qubit candidate for ultra-high-fidelity spin readout. 

\section{\label{sec_results}Results}

\begin{figure}
    \centering
    \includegraphics[width=\linewidth]{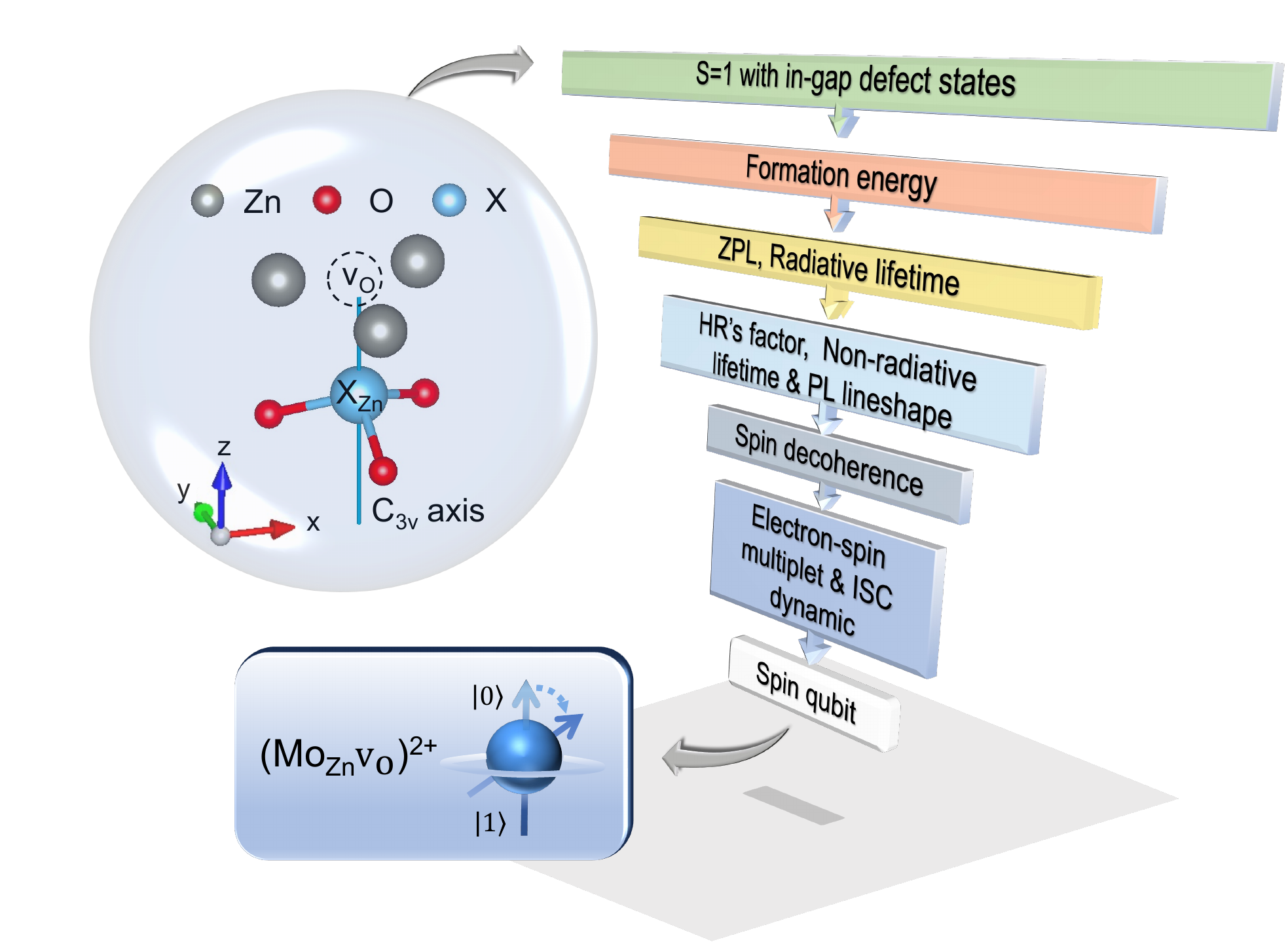}
    \caption{The computational workflow of defect candidate search with critical parameters.}
    \label{fig_structure}
\end{figure}

In Figure~\ref{fig_structure}, we propose to design an NV$^-$–like $\text{C}_{\text{3v}}$ defect complex in ZnO, consisting of an oxygen vacancy paired with a substitutional impurity (X) replacing a Zn atom~\cite{seoDesigningDefectbasedQubit2017}. To identify suitable spin qubit candidates, we follow the workflow outlined in Figure~\ref{fig_structure}, which begins with the search for defects exhibiting spin-triplet ($S=1$) ground states and in-gap defect orbitals. Thermodynamic stability is then assessed via defect formation energies and charge transition levels.

Next, we assess the optical addressability of candidate defects using two main criteria: (a) radiatively allowed in-gap optical excitations characterized by the zero-phonon line (ZPL), absorption spectrum, and radiative lifetime; and (b) optimized electron–phonon coupling, evaluated via the Huang–Rhys (HR) factor, non-radiative lifetime, quantum yield, and photoluminescence lineshape—ensuring efficient photon emission.
We further assess spin decoherence arising from hyperfine interactions with nuclear spins and other paramagnetic defects, which ultimately determine the operational duration of the qubit for sensing and computational tasks.

Finally, we evaluate the electronic–spin multiplet structure and excited-state dynamics using a combination of multireference excited-state calculations, spin–orbit coupling (SOC), zero-field splitting (ZFS), and intersystem crossing (ISC) rate predictions. Based on all these properties, we propose a promising optical readout protocol for spin defect qubits.

\subsection{Defect Formation Energy\label{sec.II.A}}
We computationally screen substitutional defects at Zn accompanied by an oxygen vacancy ($\text{v}_\text{O}$) across the periodic table, 
focusing on group B elements from groups III to VI and group A elements from groups I to V, guided by electronegativity and electron counting considerations~\cite{seoDesigningDefectbasedQubit2017}. For this purpose, we employ the hybrid density functional theory (DFT) at the HSE level of theory, which is robust for predicting ground-state properties.
This leads to the identification of four triplet defect candidates: $(\xvo)^{q}$, where $q$ is the charge state (zero or positive for n-type defects), specifically $\tivo$, $\vvo$, $\movo$, and $\nbvo$.

\begin{figure}
    \centering
        \includegraphics[width=0.9\linewidth]{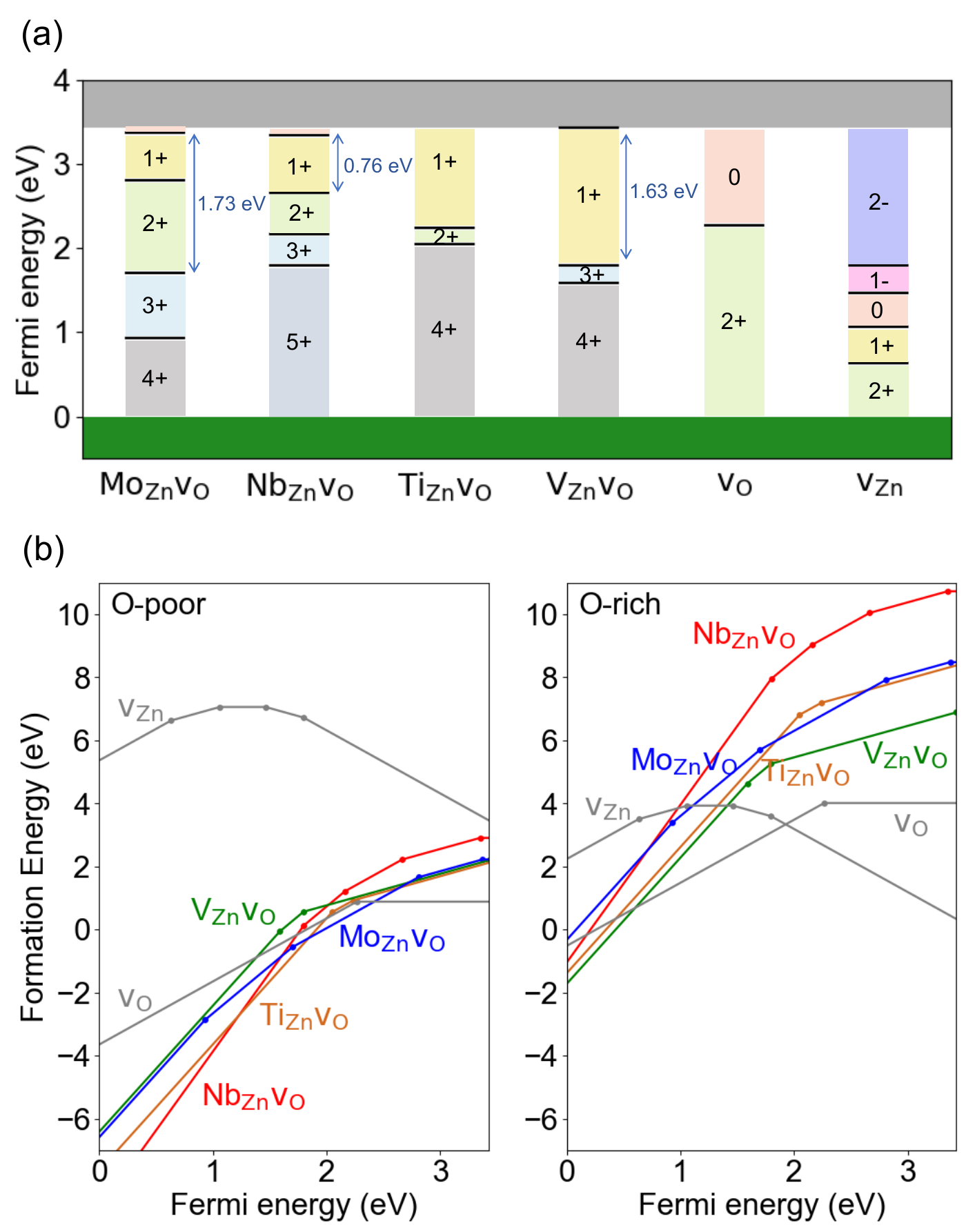}
        \caption{Defect formation energies of complex vacancy defects. (a) Charge transition levels of the complex vacancy defects  $\xvo$ (X=Mo, Nb, Ti, and V) and intrinsic defects $\text{v}_\text{O}$ and $\text{v}_\text{Zn}$ and (b) their formation energies at O-poor(left)/O-rich (right) conditions. The green/gray area in (a) corresponds to the valence/conduction band of ZnO, where VBM is set at 0 eV and the band gap is 3.43 eV.}
        \label{fig3}
\end{figure}

Figure~\ref{fig3}(a) shows the defect charge transition levels (CTL) for the 4 transition-metal-vacancy complex defects, and intrinsic vacancies ($\text{v}_\text{O}$ and $\text{v}_\text{Zn}$). The 2+, 1+, and 1+ charge states, corresponding to the triplet ground state of $\text{Mo}_{\text{Zn}}\text{v}_\text{O}$, $\text{Nb}_{\text{Zn}}\text{v}_\text{O}$, and $\text{V}_{\text{Zn}}\text{v}_\text{O}$, respectively, are stable within the gap. These defects exhibit stable positive charge states at Fermi levels up to 0.2~eV below the conduction band minimum (CBM), primarily acting as electron donors.

The charge transition level $\varepsilon$(2+/1+) for $\text{Nb}_\text{Zn} \text{v}_\text{O}$ and $\varepsilon$(3+/1+) for $\text{V}_\text{Zn} \text{v}_\text{O}$ occurs at 0.76 eV and 1.63 eV below the CBM at 3.43 eV, respectively, while $\varepsilon$(3+/2+) of $\movo$ is 1.73 eV below the CBM, classifying them as deep donor defects in ZnO. In contrast, $\tivo$ has a $\epsilon$(+1/0) transition at 0.85 eV above the CBM, which makes it unlikely to form in the neutral state. We therefore exclude $\tivo$ from further characterization.

As a benchmark, we also computed the CTL of the intrinsic defects. For $\text{v}_\text{O}$, we find the charge transition level $\varepsilon$(2+/0) to be 2.26 eV above the valence band maximum (VBM). Our calculations also show an unstable 1+ charge state, indicating a negative-$U$ center, consistent with previously reported results using hybrid functionals \cite{obaPointDefectsZnO2011, frodasonZnVacPolaron, clarkIntrinsicDefectsZnOHydrid, janottiNativePointDefects2007}. For $\text{v}_\text{Zn}$, we observe that it is an amphoteric defect with stable charge states $2-$, $1-$, $1+$, and $2+$, in agreement with previous studies~\cite{frodasonZnVacPolaron}.

Figure~\ref{fig3}(b) shows the formation energies of the aforementioned defects under O-poor and O-rich conditions.
The defect formation energies of all candidates are similar to the intrinsic donor $\text{v}_{\text{O}}$. In the n-type region (where the Fermi level close to CBM), the proposed defect formation energies are small under O-poor conditions and moderately large under O-rich conditions, compared to other intrinsic defects. In particular, intrinsic defects with smaller formation energies under O-rich conditions are mostly acceptors ($\text{v}_{\text{Zn}}$), which could compensate for the n-type donor defects. Additionally, 
we computed the formation energies of the isolated substitutional defect $\text{X}_\text{Zn}$, as shown in Supplemental Material Figure S1~\cite{SI}.

We estimate the formation probabilities of $\xvo$ complexes in proportion to their isolated components $\text{X}_\text{Zn} + \vo$ 
based on the Boltzmann distribution at a representative synthesis temperature of 1200 K.  In SI Figure~S2~\cite{SI}, we show the complex vacancy formation probabilities in the oxygen vacancy-driven n-type Fermi-level range at 1200 K ~\cite{mccluskeyDefectsZnO2009, LishuVOntype,Tsukazaki_2008} ($E_f$ between 2.26 and 3.43 eV above VBM, with CBM at 3.43 eV).

We find that the concentration of $\text{Mo}_{\text{Zn}}\text{v}_\text{O}$ increases from $0.97\%$ with $E_f$ at 2.26 eV to $3.96\%$ with $E_f$ at 3.43 eV.
We note that at synthesis temperature, e.g. 1200K, oxygen vacancies can possibly overcome the $\sim$ 1 eV activation barrier and diffuse to bind with substitutional defects $\text{X}_\text{Zn}$~\cite{LishuVOntype} to form $\text{X}_\text{Zn}\vo$. 

Once the systems' temperature is lowered to the operating temperature (i.e. room temperature), the high activation barrier will lead to an extremely small kinetic rate and lock the defect complex to its  thermodynamic equilibrium distribution at the synthesis temperature calculated above.

\subsection{Single-particle defect levels}
\begin{figure*}
    \centering
    \includegraphics[width=\linewidth]{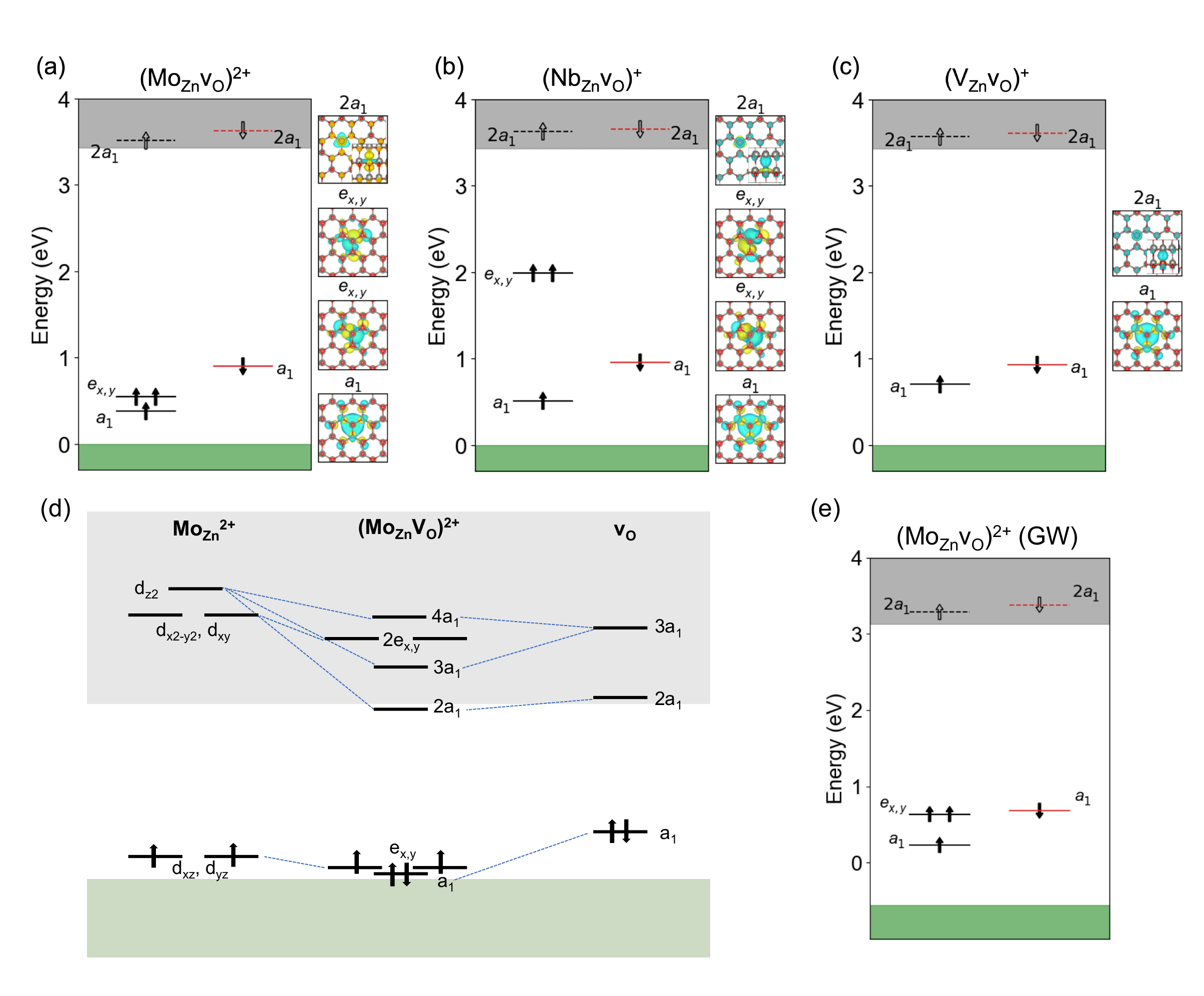}
    \caption{Single-particle levels and wavefunctions. Single-particle defect levels (horizontal black (majority spin) and red lines (minority spin)) of the (a) $\movo$, (b) $\nbvo$, (c) $\vvo$ and (e) $\vo$ defects in ZnO, calculated at the HSE ($\alpha$=0.375). The green/gray area corresponds to the valence/conduction band of ZnO, respectively. The orbitals labeled with the same symmetry symbol across three defects share the same feature and wave function natures. (d) is the schematic diagram showing the origin of defect states of $\movo$ from $\vo$ and $\mozn$, respectively. (e) Quasiparticle bandstructure at G$_0$W$_0$~\cite{govoniLargeScaleGW2015} with the PBE0 starting point for $\movo$. }
    \label{fig2}
\end{figure*}
The single-particle defect levels and wave functions of the three candidates referenced to bulk ZnO band edges are shown in Figure~\ref{fig2}(a)-(c), which provide insights of defect energy levels and molecular orbitals. To cross-check the results, we also performed the GW~\cite{govoniLargeScaleGW2015} calculations, which describe more accurately quasi-particle band structure of solids, but are computationally much more demanding than the HSE calculations. We observed consistent results between HSE and GW calculations for the spin majority channel defect level, within 0.1 eV, as shown in Figure~\ref{fig2}(e), which support the use of HSE for this system's single-particle defect levels. More numerical details can be found in Supplemental Material Section V~\cite{SI}.

All three candidates have similar defect-related orbitals, as shown in the schematic diagram of Figure~\ref{fig2}(d), which use the $\movo$ as an example. The $a_1$ and $2a_1$ states arise from the $\vo$-related states with a small mixing from the $d_z$ orbital of substitutional transition metals. The doubly degenerate $e_{x,y}$ states come from the $d_{xz}$ and $d_{yz}$ orbitals of the transition metal, hybridized with the Zn dangling bond from $\vo$ (more details for $\vo$ and metal substitutional defects can be found in Supplemental Material Section II~\cite{SI}).

The $e_{x,y}$ orbitals in the spin-majority channel lie within the band gap of ZnO for $\movo$ and $\nbvo$, while their counterparts in the spin-minority channel, and those for $\vvo$ in both channels, are more than 1.5 eV above the conduction band (CB). This results from Hund's Rules, where the $d$ orbital fillings favor open-shell configurations, causing the occupied $e_{x,y}$ orbitals in the spin-majority channel to have lower in-gap energy levels. We note that the oxidation state of the transition metal in the complex defect differs from the overall charge state \( q \) of the complex. Based on electron counting and orbital analysis, the oxidation state of the transition metal ions are identified as \( \text{Mo}^{+4} \), \( \text{Nb}^{+3} \), and \( \text{V}^{+3} \) in their respective complex structures, all corresponding to a $d^2$ electron configuration.

\begin{table*}[!ht]
\caption{\label{tab1} Summary of the optical properties of the proposed defects in their triplet states. The ZPL values listed represent the finalized choices obtained from the selected levels of theory,
with DLPNO-NEVPT2@CASSCF used for the spin-majority transition of $\movo$
and mcDFT used for all remaining transitions.
Also reported are the relaxation energy computed with mcDFT ($E_{\text{rel}}$), 
the squared transition dipole moment ($\mu^2$), 
and the radiative lifetime ($\tau_R$) for each ZnO defect system.}
\renewcommand{\arraystretch}{1.2}
\begin{ruledtabular}
\begin{tabular}{cccccc}

Spin &Transition &$\text{ZPL}$(eV) &$E_{\text{rel}}$(eV) & $\mu^2$ (bohr$^2$) & $\tau_R$ ($\mu$s) \\
 \hline
 \multicolumn{6}{c}{$\movo$} \\
\hline
 Maj &e $\ra$ $\aone$ (x,y) &2.08 &0.18 & $4.96 \times 10^{-2}$ & 5.03 \\
 Min &a$_1$ $\ra$ 2$\aone$ (z) &1.56 &0.85 & $2.20 \times 10^{-2}$ & 27.10 \\
 \hline
 \multicolumn{6}{c}{$\nbvo$} \\
 \hline
 Maj &e $\ra$ $\aone$ (x,y) &1.07 &0.51  & $8.09 \times 10^{-2}$ &36.8 \\
 Min &a$_1$ $\ra$ 2$\aone$ (z) &1.72 &0.74 & $2.82 \times 10^{-2}$ & 15.7 \\
 \hline

 \multicolumn{6}{c}{$\vvo$} \\
 \hline
 Maj &a$_1$ $\ra$ 2$\aone$ (z) &0.13 &2.49 & $6.09 \times 10^{-2}$ & 1.69$\times 10^{4}$  \\
 Min &a$_1$ $\ra$ 2$\aone$ (z) &1.62 &0.86 & $2.19 \times 10^{-3}$ & 244.00 \\

\end{tabular}
\end{ruledtabular}
\end{table*}

\begin{table*}[!ht]
    \centering
    \caption{\label{E_es} Computed ZPL energies (eV) of $\movo$ in ZnO at different levels of theory.
    }
    \renewcommand{\arraystretch}{1.2}
        \begin{ruledtabular}
    \begin{tabular}{l|cccc}
        \multirow{2}{*}
{\textbf{Theory}} & \multicolumn{4}{c}{\textbf{ZPL Energies (eV)}\footnotemark[1]} \\

        &$^3A_2$& $^1E$ & $^1A_1$ & $^3E$ \\
\hline
        mcDFT@HSE &0& 0.83 \footnotemark[2] & -- & 1.99 \footnotemark[3]  \\
         QDET(14e14o) & 0 & -- & -- & 1.88 \\
        DLPNO-NEVPT2@CASSCF(4,7) &0 &  0.81 & 1.63 \footnotemark[4] & 2.08 \\
    \end{tabular}

\begin{minipage}{\textwidth}
\footnotetext[1]{For each theory, the ZPL energy is calculated as the vertical excitation energy (with a theory stated on the left column) at the fixed ground-state geometry minus the excited-state relaxation energy calculated from mcDFT. }
\footnotetext[2]{The relaxation energy of $^1E$ from mcDFT is 0.12 eV. }
\footnotetext[3]{The relaxation energy of $^3E$ from mcDFT is 0.18 eV. }
\footnotetext[4]{The geometry of $^1A_1$ is assumed to be the same as $^3A_2$ due to the similar non-degenerate symmetry.}
\end{minipage}
    \end{ruledtabular} 
\end{table*}

\subsection{ZPL, absorption spectra and radiative lifetime}
In this section, we begin our investigation of the spin-photon interface by determining the dipole moments, zero-phonon line (ZPL) energies, and radiative lifetimes ($\tau_R$) of the defect candidates. These quantities provide the key inputs for estimating the brightness and spin-dependent optical contrast of the defect, which in turn determine the achievable fidelity of optical initialization and readout.

First, we compute the absorption spectra of the triplet states within the Random-Phase Approximation (RPA) using single particle wavefunctions computed at HSE (Figure~S5 in the Supplemental Material~\cite{SI}). Local-field effects~\cite{onidaElectronicExcitationsDensityfunctional2002} were included, as they are crucial in nonhomogeneous dielectric environments such as defect systems. The defect transition energies are well separated from the bulk excitation, spanning from ultra-red to blue. The squared dipole moments ($\mu^2$) for defect transitions, for light polarized along the $x$, $y$, and $z$ directions, were extracted from the spectra. These values are summarized in Table~\ref{tab1}, where the relevant polarization direction for each non-negligible $\mu^2$ is indicated by “$x$/$y$/$z$” in the transition column. We note that these computed dipole moments are consistent with the group-theory analysis for $C_{3v}$, where the e $\ra$ 2$\aone$ transition is allowed only when light is polarized in the $x$-$y$ plane for the majority spin channel, while the $\aone$ $\ra$ 2$\aone$ transition is allowed only when light is polarized along the $z$ direction for the minority spin channel. 

We next evaluate the zero-phonon line (ZPL) energies, defined as the energy difference between the minima of the excited- and ground-state potential surfaces. For accurate prediction and cross-validation, we employed three different computational methods: (i) multideterminant constrained DFT (mcDFT; details in Section~\ref{sec-mcDFT}), (ii) Quantum defect embedding theory (QDET)~\cite{maFirstprinciplesStudiesStrongly2020,maQuantumSimulationsMaterials2020,shengGreensFunctionFormulation2022,vorwerkQuantumEmbeddingTheories2022}, and (iii) the multiconfigurational self-consistent field method with perturbative correction (DLPNO-NEVPT2@CASSCF)~\cite{guoSparseMapsSystematicInfrastructure2016} (details in Supplemental Material Sections V and XI~\cite{SI}). The mcDFT method, while mean-field in nature, provides a practical balance between accuracy and computational cost, whereas QDET and DLPNO-NEVPT2@CASSCF are higher-level approaches better suited for strongly correlated open-shell excited states of transition-metal defects in oxides. 

Table~\ref{E_es} summarizes the energies of the triplet ground state $\gs$, the triplet excited state ($\tes$), and the singlets ($\sgs$ and $\fses$) for $\movo$, where the doubly- degenerated many-body $^3E$ state is characterized by the $e \rightarrow a_1$ transition, and the detail of single particle configuration for states can be found in Section~\ref{sec-mcDFT} and Table S11~\cite{SI}. The three approaches yield consistent results for the $\tes$ triplet excited state and the $\sgs$ singlet, confirming that mcDFT can reliably capture the excited-state properties of this system. A more detailed comparison of the optical properties across methods is provided in Supplemental Material Section XIV~\cite{SI}. For the remainder of our study, we adopt the DLPNO-NEVPT2@CASSCF result as the ZPL reference for our key defect candidate, $\movo$, while using mcDFT values for other defects as a balance between accuracy and computational efficiency. The final ZPL results are summarized in Table~\ref{tab1}, and these values are used throughout the remainder of this work whenever a ZPL input is required.

Based on the ZPL energy and transition dipole moment, we calculate radiative lifetimes ($\tau_R$) scales as $E^{-3}\mu^{-2}$ (Eq.~\ref{equation_rad}) for $\movo$ and $\nbvo$, as summarized in Table~\ref{tab1} (details in Supplemental Material Section IV~\cite{SI}).  For example, the $e \rightarrow 2a_1$ transition in $\movo$ has the shortest lifetime ($\thicksim$ 5.03 $\mu s$).
Comparing with other spin defect qubits, the radiative lifetimes of the proposed defects are longer than the NV center (12 ns)~\cite{maExcitedStatesNegatively2010a}, but shorter than the boron vacancy in hBN (20 $\mu$s)~\cite{ivadyInitioTheoryNegatively2020}. The difference is mainly due to variations in dipole moments based on Eq.~\ref{equation_rad}, i.e. the NV center has a transition dipole moment ($\mu^2 \sim 3.87  
 a.u.$~\cite{liExcitedstateDynamicsOptically2024a}), much larger than that of our proposed systems ($\mu^2 \sim 10^{-2} a.u.$), while the boron vacancy exhibits a much smaller one ($\mu^2 \sim 10^{-4} a.u.$~\cite{reimersPhotoluminescencePhotophysicsPhotochemistry2020}).

\subsection{Huang-Rhys factor, Non-radiative recombination and Photoluminescence lineshape}
\begin{figure*}
    \centering
    \includegraphics[width=\linewidth]{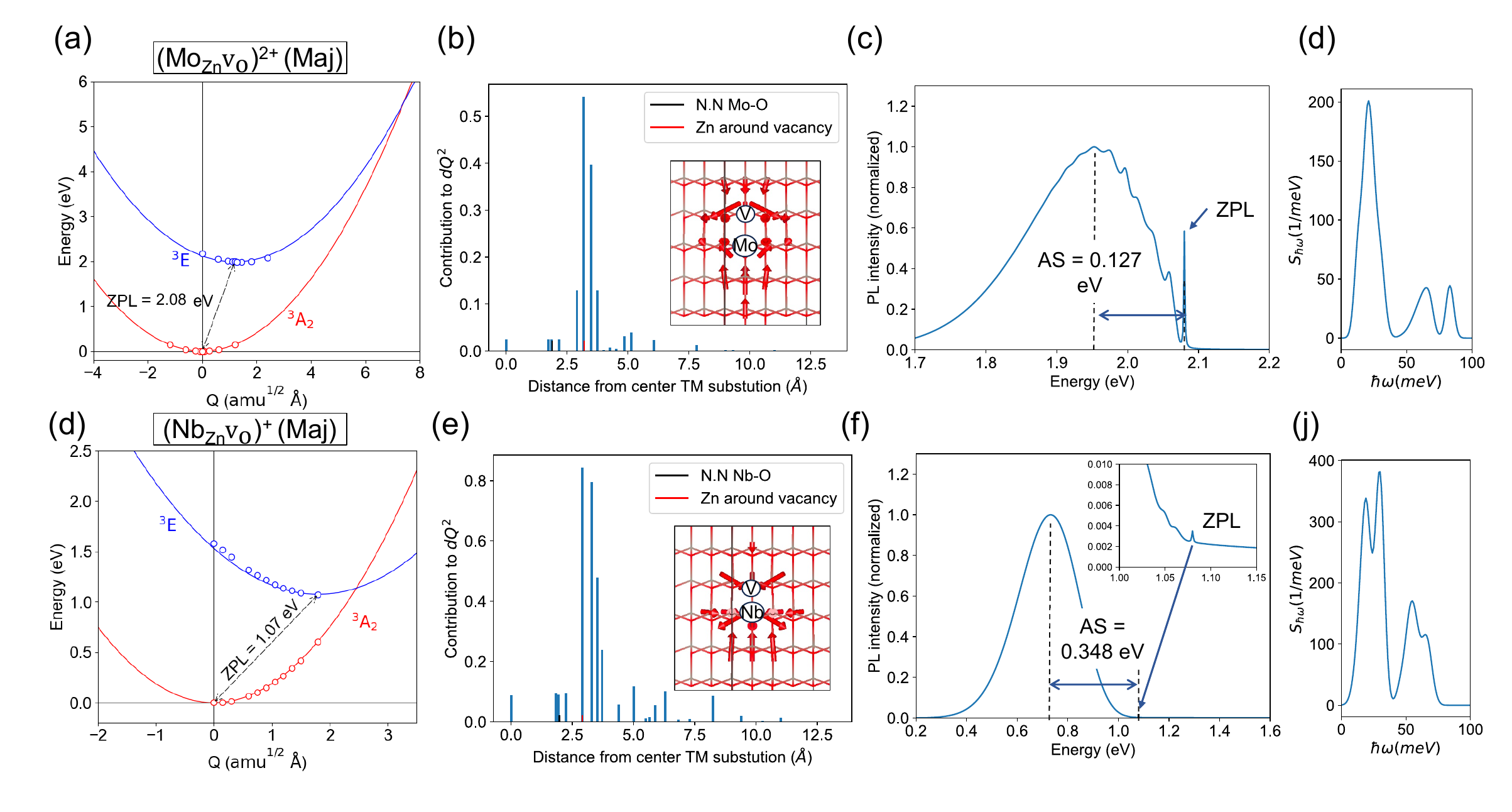}
    \caption{\label{Fig567}Electron-phonon properties of $\movo$ and $\nbvo$ spin majority transition between triplet states. The panel (a-d) are for $e \rightarrow a_1$ transition in $\movo$, and (d-g) are for $e \rightarrow a_1$ transition in $\nbvo$. (a,d) are the configuration coordinate diagram of the potential surface (dots) fitted by harmonic potentials with effective phonon modes(lines). (b,e) are the module square of mass weighted displacement $dQ^2$ distribution for proposed defects contributed from atoms as a function of radius from the TM defect center. The contribution to $dQ^2$ from atom at position $r$ from the TM site, is defined as $dQ^2 (r) = \sum_\alpha dQ^2(\alpha)|_{|r_{\alpha}-r|<0.1}$, where the $r_\alpha$ is the atom $\alpha$'s distance from TM site and 0.1 is the step size of the sampling. The total displacement $\Delta Q=\int dQ^2(|\vec{r}|) d\vec{r}$. (c,f) are the photoluminescence(PL) lineshape, where the anti-stoke shifts(AS) are determined by the energy difference between the highest peak of phonon side band (PSB) and the ZPL. (d,j) are the corresponding phonon spectrum functions revealing the partial HR factor contributed by the phonon mode with energy $\hbar\omega$. }
\end{figure*}
\begin{table*}[!ht]
\caption{\label{tab_nonrad} The phonon-related properties of optical spectroscopy for $\movo$ and $\nbvo$. $\Delta Q$ is the mass-weighted displacements between the excited and ground state geometries. $k\approx\frac{\text{ZPL}}{\hbar\omega_{eff}\text{($gs$)}}$ is used to approximate the number of phonons necessary to mediate the transition. $S_f^{\text{eff}}$ is the HR's factor computed by effective phonons for ground state. $\tau_{NR}$ is the non-radiative lifetime. $W_{if}$ and $X_{if}$ are the electronic part and phonon part of the nonradiative lifetime ($\tau_{NR}$). QY is the quantum yield. 
}
\renewcommand{\arraystretch}{1.2}
\begin{ruledtabular}
\begin{tabular}{cccccccccc}
 Spin &Transition & ZPL &$\Delta Q$ & k & $\tau_{NR}$ & $W_{if}$ & $X_{if}$ & $S_f^{\text{eff}}$ & QY \\
  & & (eV) & (amu$^{1/2}$Å) & & ($\mu$s) & ($\text{eV}/(\text{amu}^{1/2}\text{\AA}$)) & &  & \\
 \hline
 \multicolumn{10}{c}{$\movo$} \\
 \hline
 Maj &e $\ra 1\aone$  (x,y)&2.08 &1.19  & 72 & Not allowed & $5.13\times10^{-2}$ & $1.24\times10^{-29}$ & 4.96&High ($\approx1$)\\
 Min &$1\aone \ra 2\aone$ (z)&1.56 &5.05 & 91 & $1.07\times10^{-7}$ & $3.75\times10^{-2}$  & $1.23\times10^{-1}$ & 50.64 &Small ($\approx0$)\\
 \hline
\multicolumn{10}{c}{$\nbvo$} \\

 \hline
 Maj &e $\ra 1\aone$ (x,y)&1.07 &1.80 & 31 & $1.63\times10^{-9}$ & $4.42\times10^{-2}$ & 5.82 & 15.36 &Small ($\approx0$) \\ 
 Min &$1\aone \ra 2\aone$ (z)&1.72 &3.83 & 84 & $2.47\times10^{-5}$ & $4.62\times10^{-2}$ & $3.52\times10^{-4}$ & 35.67 &Small ($\approx0$) \\ 
\end{tabular}
\end{ruledtabular}
\end{table*}

To further characterize the phonon-related optical properties of the defect candidates, we compute the Huang-Rhys (HR) 
factors~\cite{markhamInteractionNormalModes1959,walkerOpticalAbsorptionLuminescence1979}, the non-radiative lifetime ($\tau_{NR}$)~\cite{wuCarrierRecombinationMechanism2019}, as well as the quantum yield, as summarized in Table \ref{tab_nonrad}.
The HR factor quantifies the strength of electron-phonon coupling and the energy reorganization during electronic transitions. Small electron-phonon coupling strength gives a small HR factor, which leads to the relative high intensity of the ZPL compared to the phonon sideband (PSB). Non-radiative ($NR$) recombination refers to the electron-hole recombination mediated by phonons, which does not contribute to photoluminescence. The optical brightness, determined by the competition between the radiative and $NR$ recombination paths, is characterized by the quantum yield (QY = $\frac{1/\tau_R}{1/\tau_R + 1/\tau_{NR}}$).

Among the spin defects we studied in Figure~\ref{fig2}, all the $a_1 \rightarrow 2a_1$ transitions show large HR factors and fast non-radiative decay, leading to a large PSB and low quantum yield. This is consistent with the excited state relaxation energy ($E_{\text{rel}}$) prediction in Table~\ref{tab1}, where the $E_{\text{rel}}$ for the $a_1 \rightarrow 2a_1$ transition are generally larger than those for the $e \rightarrow a_1$ transition, indicating stronger electron-phonon coupling. 
From the orbital perspective, the $a_1 \rightarrow 2a_1$ transition in all three systems are mainly originated from the internal transition of $\vo$ defect, with small mixing from the $d_z$ orbital of the transition metal (TM) ion, as shown in the Figure~\ref{fig2}(d), which exhibits significant electron-phonon coupling and a large lattice relaxation energy ($\thicksim 1$  eV). We also note that this large electron-phonon coupling associated with the Zn dangling bonds are consistent with previous study~\cite{lyonsFirstprinciplesCharacterizationNativedefectrelated2017a}. Based on these considerations, we exclude all spin-minority transitions as well as the $\vvo$ defect from the list of viable candidates.

The spin-majority transitions of $\movo$ and $\nbvo$ have smaller HR factors (4.96 and 15.36, respectively) compared to the spin-minority channels. However, $\nbvo$ still shows a fast non-radiative decay and low quantum yield. In contrast, $\movo$ combines a small HR factor with slow non-radiative decay, making it the best candidate.

To gain insights on determining factors of HR factors, we compare the two proposed defects with others in ZnO and various host materials in Table~\ref{tab_HR}. We observe that host materials with light elements (C, BN) generally exhibit smaller HR factors compared to ZnO or GaN, which can be attributed to the dependence of atomic mass ($m_{\alpha}$) in $\Delta Q$, hence HR factor $S_f^{eff}$ ($S_f^{eff}=\omega_{eff}\Delta Q^2/2\hbar$ and  $\Delta Q=\sqrt{\sum_{\alpha,t}m_{\alpha}\Delta R_{\alpha,t}^2}$~\cite{markhamInteractionNormalModes1959}). Furthermore, the dominant phonon frequency ($\omega_{eff}$) varies within a factor of 2, particularly for defects in ZnO. Thus, the primary factor influencing HR variation within a given host (e.g., ZnO) is the displacement between excited and ground states ($\Delta R_{\alpha,t}$). 
\begin{table}[!ht]
\caption{\label{tab_HR}A summary of HR factor (S) of various defects from past literature and this work. }
\renewcommand{\arraystretch}{1.0}
\begin{ruledtabular}
\begin{tabular}{cccc}
Defect(host)                & $\Delta Q$(amu$^{1/2}\AA$)           & $\hbar\omega$(meV)     & S \\
\hline
$\movo$(ZnO)  &1.19                                  &29                         &4.96\\
$\nbvo$(ZnO)  &1.80                                  &40                         &15.36\\
$\text{Li}_{\text{Zn}}$~\cite{alkauskasFirstprinciplesTheoryNonradiative2014b}(ZnO)                     
       & 3.22                                 & 36                        &28\\
$\text{N}_{\text{O}}$~\cite{alkauskasFirstPrinciplesCalculationsLuminescence2012a}(ZnO) 
       & 1.92                                 & 40                        &15.3\\
$\text{Cu}_{\text{Zn}}^{-\ra0}$~\cite{lyonsDeepDonorState2017}(ZnO) 
       & $\approx $1.39\footnotemark[1]       & 57                        &13.1\\
$\text{Cu}_{\text{Zn}}^{+\ra0}$~\cite{lyonsDeepDonorState2017}(ZnO) 
       & $\approx $1.33\footnotemark[1]       & 52                        &11.0\\
$\text{$\text{V}_\text{Zn}\text{Al}_\text{Zn}^{0\ra-}$}_{\text{ }}$~\cite{frodasonZnVacancydonorImpurity2018}(ZnO) 
       & 2.7                                 & 33                        &25\\
$\text{$\text{V}_\text{Zn}\text{Si}_\text{Zn}^{+\ra0}$}_{\text{ }}$~\cite{frodasonZnVacancydonorImpurity2018}(ZnO) 
       & 2.7                                 & 33                        &25\\
$\text{$\text{V}_\text{Zn}\text{H}$}^{0\ra-}$~\cite{frodasonZnVacancydonorImpurity2018}(ZnO) 
       & 2.95                                 & 30                        &27\\

$\text{C}_\text{N}$~\cite{alkauskasFirstprinciplesTheoryNonradiative2014b}(GaN)
       & 1.61                                 & 42                        &10\\
 $\text{Zn}_{\text{Ga}}\text{V}_{\text{N}}$~\cite{alkauskasFirstprinciplesTheoryNonradiative2014b}(GaN)                       
       & 3.33                                 & 26                        &30\\
NV$^-$~\cite{alkauskasFirstprinciplesTheoryLuminescence2014}(Diamond) 
       & $\approx $0.7\footnotemark[1]       & 64                      & 3.67 \\
Cl$-$V~\cite{bulancea-lindvallChlorineVacancySiC2023}(4H-SiC)
       & $\approx $0.96\footnotemark[1]       & 32 to 40                & 3.5 to 4.4\\
VB$^-$~\cite{ivadyInitioTheoryNegatively2020,reimersPhotoluminescencePhotophysicsPhotochemistry2020}(hBN) 
       & $\approx $1.1\footnotemark[1]       & 24.8                    &3.69\\
$\text{C}_\text{B}\text{C}_\text{N}$~\cite{mackoit-sinkevicieneCarbonDimerDefect2019}(hBN) 
       & $\approx $0.38\footnotemark[1]       & 120                     &2\\
$\text{N}_\text{{B}}\text{V}_\text{N}$~\cite{tawfikFirstprinciplesInvestigationQuantum2017}(hBN) 
       & 0.66                                 & 30,46,48 \footnotemark[2] &4.49\\

\end{tabular}
\footnotetext[1]{$\Delta Q$ is estimated by S, through $S=\omega_{eff}\Delta Q^2/2\hbar$.}
\footnotetext[2]{phonon modes which have the largest contribution to the HR factor; the rest of $\hbar\omega$ in the table are effective phonon frequencies.}
\end{ruledtabular}
\end{table}
Our proposed $\movo$ defect (spin-majority channel $e$ $\rightarrow$ $1\aone$ transition) exhibits a surprisingly small HR factor of 4.96, which, while slightly larger than that of the NV center in diamond, remains one of the smallest among known defects in ZnO. In comparison, experiments have successfully resolved the zero-phonon line (ZPL) of copper substitutional defects in ZnO, which have an HR factor of 11~\cite{lyonsDeepDonorState2017}. This suggests that the ZPL of $\movo$ should also be resolvable experimentally.

To understand why $\movo$ defect has a smaller HR factor than other defects in ZnO, we compare $\movo$ with $\nbvo$ as an example. 
In Figures~\ref{Fig567}(d,e), we show the effective phonon mode and the contribution to $\Delta Q^2$ from atoms in real space for the spin majority transition of $\movo$ and $\nbvo$. Both of the two systems have an effective phonon mode
localized within 5 $\text{\AA}$ from the TM defect center. The major contributors to $\Delta Q^2$ arise from the Zn atoms around the vacancy and next-nearest neighbors of the TM substitution site. With similar atomic mass contributions (single atom Mo or Nb mass difference is negligible in the summation in $\Delta Q$), the smaller $\Delta Q^2$ of $\movo$ is mostly from overall smaller atomic displacement ($\Delta R_{\alpha}$) between ground and excited states. This difference between two defects may arise from the closer ionic radius between Mo (oxidation state +4, coordination VI, 0.65~\AA) and Zn (oxidation state +2, coordination IV, 0.60~\AA), compared to Nb (oxidation state +3, coordination VI, 0.72~\AA)~\cite{shannonRevisedEffectiveIonic1976}. 

We next discuss photoluminescence (PL) lineshape and phonon sidebands of the $\movo$ and $\nbvo$ spin-majority transitions using the generating function approach with all phonons~\cite{alkauskasFirstPrinciplesCalculationsLuminescence2012a}. As shown in Figure \ref{Fig567}(c,f), the $\movo$ exhibits a sharp and high intensity ZPL peak and a phonon side band with an anti-Stokes (AS) shift of 0.127 eV. On the other hand, the ZPL of $\nbvo$ is relatively small, but its AS shift of 0.348 eV may allow it to be distinguishable from the large PSB.
The corresponding partial HR factor spectrum functions are shown in Figure~\ref{Fig567}(d,j). We found $\movo$ has a dominant peak at 21 meV, and $\nbvo$ between 19-30 meV and 55-65 meV. To understand the localization of these phonon modes, we compute the inverse participation ratio (IPR), estimating the number of atoms involved in the vibration of the phonon modes~\cite{liCarbonTrimerEV2022}. We found the IPRs are all larger than 20 for the two defects, representing delocalized bulk-like phonon modes. Their energies lie within the acoustic band of bulk ZnO.

In the Franck-Condon picture, the AS shift corresponds to the ground state energy difference between its equilibrium geometry and excited-state geometry. We found their values (0.13 eV for $\movo$ and 0.35 eV for $\nbvo$) to be close to the values from CDFT relaxation energy of ground state, i.e.
0.14 for $\movo$ and 0.50 eV for $\nbvo$, respectively. In addition, the HR factor obtained by all-phonon-mode summation is in reasonable agreement with the one with an effective phonon along the configuration coordinate~\cite{wuCarrierRecombinationMechanism2019}, which validates the consistency between different theoretical approaches in this system (More details can be found in Sec. VIII of the Supplemental Material~\cite{SI})

Finally, to explain why the spin-majority non-radiative (NR) transition is forbidden in $\movo$ but fast in $\nbvo$, we refer to classical Marcus theory in the high-temperature limit, as illustrated in Figure~\ref{Fig567}(a,d). In this picture, the potential energy surface (PES) crossing between the ground state (red curve) and the excited state (blue curve) defines an energy barrier that the phonon-assisted $NR$ recombination process needs to overcome. A much larger barrier is obtained for $e \rightarrow 2a_1$  transition in $\movo$ than $\nbvo$, which explains the much slower $NR$ process in $\movo$. This could be understood from the much larger ZPL and smaller $\Delta Q$ in $\movo$. 

From a theoretical perspective, the phonon-assisted $NR$
recombination rate, within the static coupling approximation and the one-dimensional effective phonon approximation, is expressed with the product of two terms:
the square of electronic term ($W_{if}$) and the phonon term ($X_{if}$) (More details in Section~\ref{sec-Non-rad})~\cite{wuCarrierRecombinationMechanism2019}. The electronic coupling term exhibits relatively small variation between two defects as shown in Table~\ref{tab_nonrad}.
The phonon term $X_{if}$ of $\nbvo$ is many orders of magnitude larger than that of $\movo$. Consequently, $\nbvo$ exhibits stronger electron-phonon coupling and a faster $NR$ decay. More detailed discussion can be found in Supplemental Material Figure S9~\cite{SI}.

\subsection{Spin decoherence time}

\begin{figure}[h]
    \centering
    \includegraphics[width=1\linewidth]{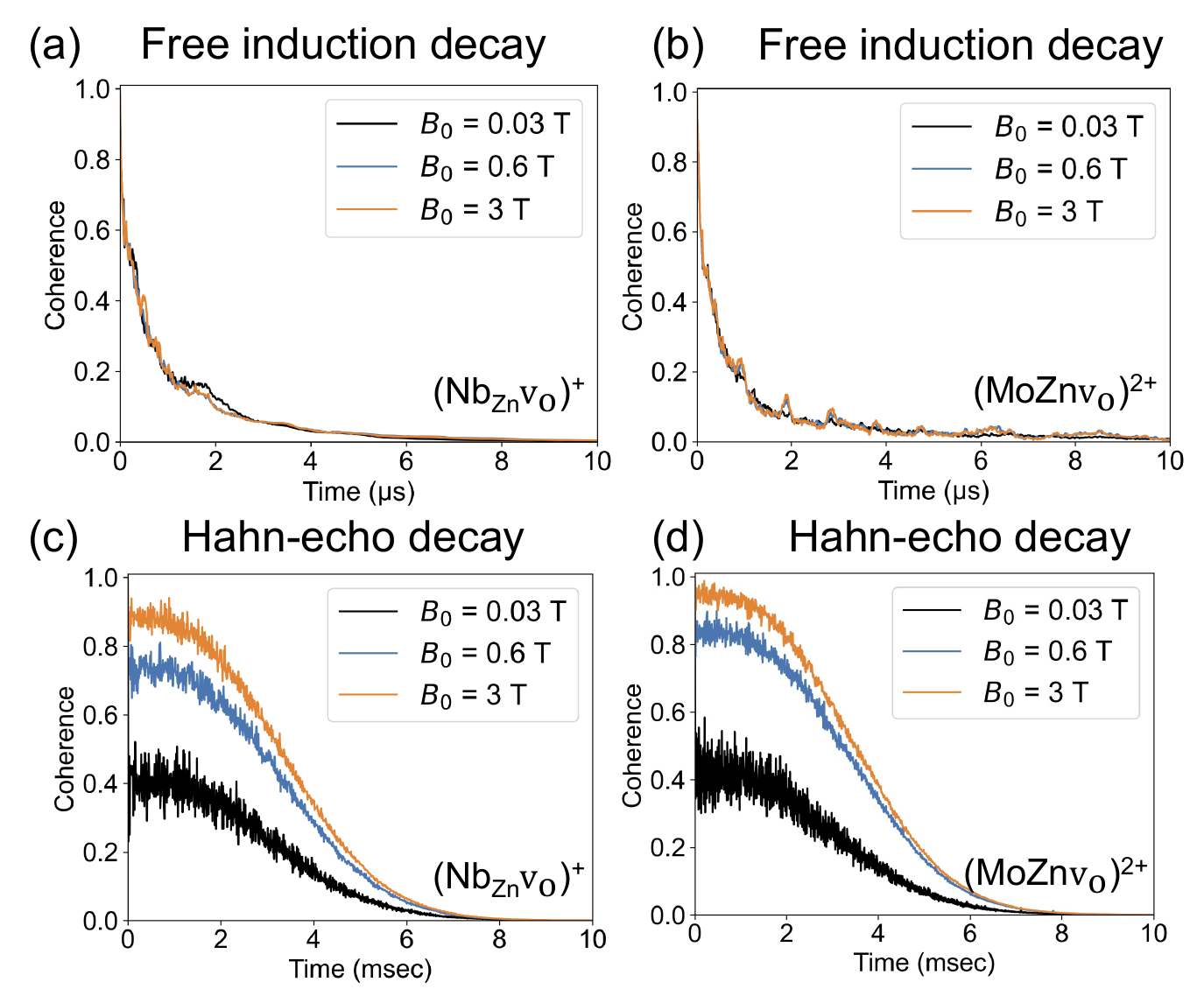}
    \caption{Spin decoherence of transition metal-vacancy complex in a nuclear spin bath ZnO. (a, b) Free induction decay ($\ttwo^*$) of $\nbvo$ (a) and $\movo$ (b) in ZnO under various external magnetic field strengths ($B_0$). (c, d) Hahn-echo coherence decay ($\ttwo$ ) of $\nbvo$ (c) and $\movo$ (d) in ZnO under various external magnetic field strengths.}
    \label{fig_t2}
\end{figure}

In order to predict spin decoherence time of qubit candidates, we compute the free-induction decay (FID) signals due to hyperfine couplings with the nearby nuclear spins of $^{67}$Zn (I=5/2, 4.1$\%$) and $^{17}$O (I=5/2, 0.038$\%$). Figure~\ref{fig_t2}(a) and (b) show the FID signals for $\nbvo$ and $\movo$, respectively, under three different external magnetic field strengths. We observe that the FID signals saturate at 300 G, with maximum decay time ($\ttwo^*$) of 0.40 $\mu s$ and 0.37 $\mu s$ for $\movo$ and $\nbvo$, respectively. Both exhibit strong non-Gaussian features with stretched exponents of 0.51 and 0.56 for $\movo$ and $\nbvo$, respectively, due to a strong anisotropic hyperfine coupling~\cite{zhaoDecoherenceDynamicalDecoupling2012}. We remark that the $\ttwo^*$ times of these qubit candidates are an order of magnitude smaller than that of diamond NV centers~\cite{mazeFreeInductionDecay2012} and of a silicon vacancy in SiC~\cite{nagyHighfidelitySpinOptical2019} due to the larger magnetic moment and abundance of $^{67}$Zn nuclear spins than those of $^{13}$C nuclear spins (1.1 $\%$) in diamond. However the $\ttwo^*$ is order of magnitude larger than the shallow donor in ZnO~\cite{linpengCoherencePropertiesShallow2018}

\begin{figure}[h]
    \centering
    \includegraphics[width=1\linewidth]{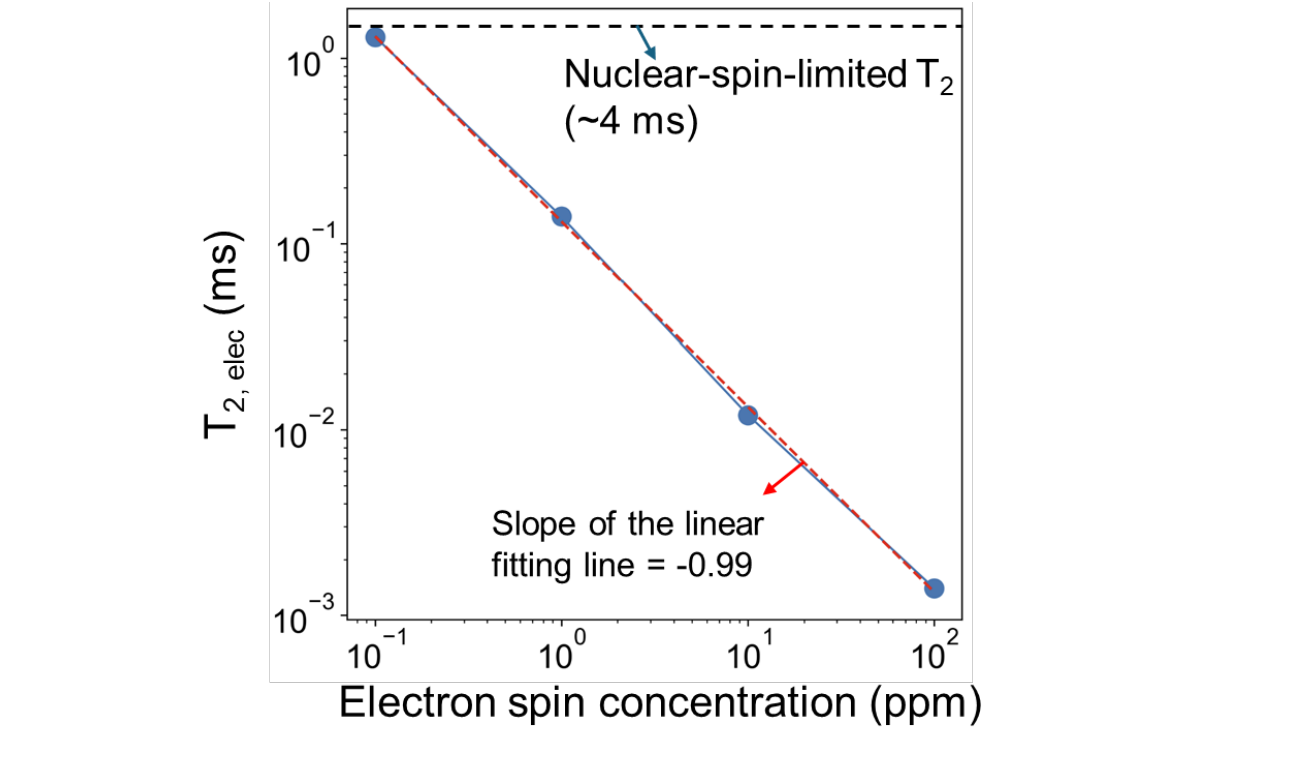}
    \caption{Electronic spin-induced decoherence time $\ttwo$ as a function of electron spin concentrations. The black dashed line shows the $\ttwo$ time determined by the nuclear spin bath in ZnO and the red dashed line represents a linear fitting model in logarithmic scale.}
    \label{fig_t2_concentration}
\end{figure}

The inhomogeneous broadening effect in FID can be removed by applying the Hahn-echo pulse sequence. Figure \ref{fig_t2}(c) and (d) show the computed Hahn-echo signals of the $\nbvo$ and $\movo$, respectively, under three different magnetic field strengths. Notably, we observe that the decoherence occurs on two different time scales: an initial partial collapse within the $\mu s$ range, followed by a gradual decay on the millisecond timescale.  The gradual coherence decay is due to the dynamical fluctuation of the nuclear spin bath arising from the magnetic dipolar coupling between nuclear spins~\cite{zhaoDecoherenceDynamicalDecoupling2012}. We find that the 
gradual decay saturates above 300 G and is described by a stretched exponential function with exponents of 2.49 (2.32) and the $\ttwo$ times of 4.04 (3.96) ms for $\movo$, and $\nbvo$ in the parentheses, respectively. This saturation is attributed to the suppression of spin-flip transitions other than the spin flip-flop transitions in the bath, caused by large energy gaps between nuclear spin levels due to the Zeeman effect~\cite{zhaoDecoherenceDynamicalDecoupling2012}.

For the early partial collapse observed in Figure \ref{fig_t2}(c) and (d), we attribute its origin to strain-induced inhomogeneous quadrupole interactions around the defect site (see SI~\cite{SI} Table S9). The variation in quadrupole interactions leads to differing precession frequencies of the nuclear spins~\cite{roseSpinCoherence14N2017}, causing irregular electron spin echo envelope modulation (ESEEM)~\cite{mimsEnvelopeModulationSpinEcho1972,yangElectronSpinDecoherence2014,bulancea-lindvallIsotopePurificationInducedReductionSpinRelaxation2023}. In SI~\cite{SI} Figure S12, we demonstrate that this early coherence collapse disappears in a hypothetical bath model, where quadrupole interactions are removed from the spin Hamiltonian, confirming that the collapse is due to quadrupole interactions. Additionally, the partial coherence collapse disappears above 3 T, where the Lamor frequency becomes the dominant frequency of the nuclear spins, suppressing the ESEEM depth.

In addition to nuclear spins, ZnO contains intrinsic defects, some of which may be paramagnetic~\cite{vlasenkoOpticalDetectionElectron2005a,zuoFerromagnetismPureWurtzite2009, fedorovInvestigationIntrinsicDefect2017, linFirstprinciplesStudyMagnetic2019}, even in isotopically purified materials. In such engineered samples, where nuclear spins are removed, qubit decoherence would be dominated by the electron spin bath from paramagnetic defects. To quantify this effect, we compute the Hahn-echo $\ttwo$ time as a function of electron spin-1/2 concentration in ZnO, ranging from 0.1 ppm to 100 ppm ($10^{16}-10^{19} \text{cm}^{-3}$), which is consistent with experimental defect concentrations~\cite{tuomistoEvidenceZnVacancy2003,savchenkoRoleParamagneticDonorlike2020}. As shown in Figure~\ref{fig_t2_concentration}, when the electron spin concentration reaches 0.035 ppm ($2.85 \times 10^{15}$ $\text{cm}^{-3}$), the $\ttwo$ time induced by the electron spin bath becomes comparable to that caused by the nuclear spin bath. Our results suggest that for isotopic purification to be effective, the paramagnetic defect concentration should be kept below 0.035 ppm. 

As the electronic spin concentration increases beyond 10$^{16}$ $\text{cm}^{-3}$, the $\ttwo$ time is rapidly reduced to sub-millisecond. We find that the $\ttwo$ time decreases almost linearly in the log scale with a slope of -0.99. We note that nuclear spins can be removed or significant reduced by nuclear spin purification, then the paramagnetic impurities would be the main source for spin decoherence. 

\subsection{Spin properties and optical read-out}

The comprehensive analysis of the thermodynamic properties, optical properties and electron-phonon coupling, as well as spin decoherence time of these transition metal vacancy complexes in ZnO demonstrated their suitability for optically-addressable spin qubits. $\movo$ stands out to be the most promising, with its low formation energy and optimal combination of other characterisitics. We therefore focus on discussing the readout of spin qubit $\movo$ through a spin-photon interface as follows.

\begin{figure}
    \centering
    \includegraphics[width=1\linewidth]{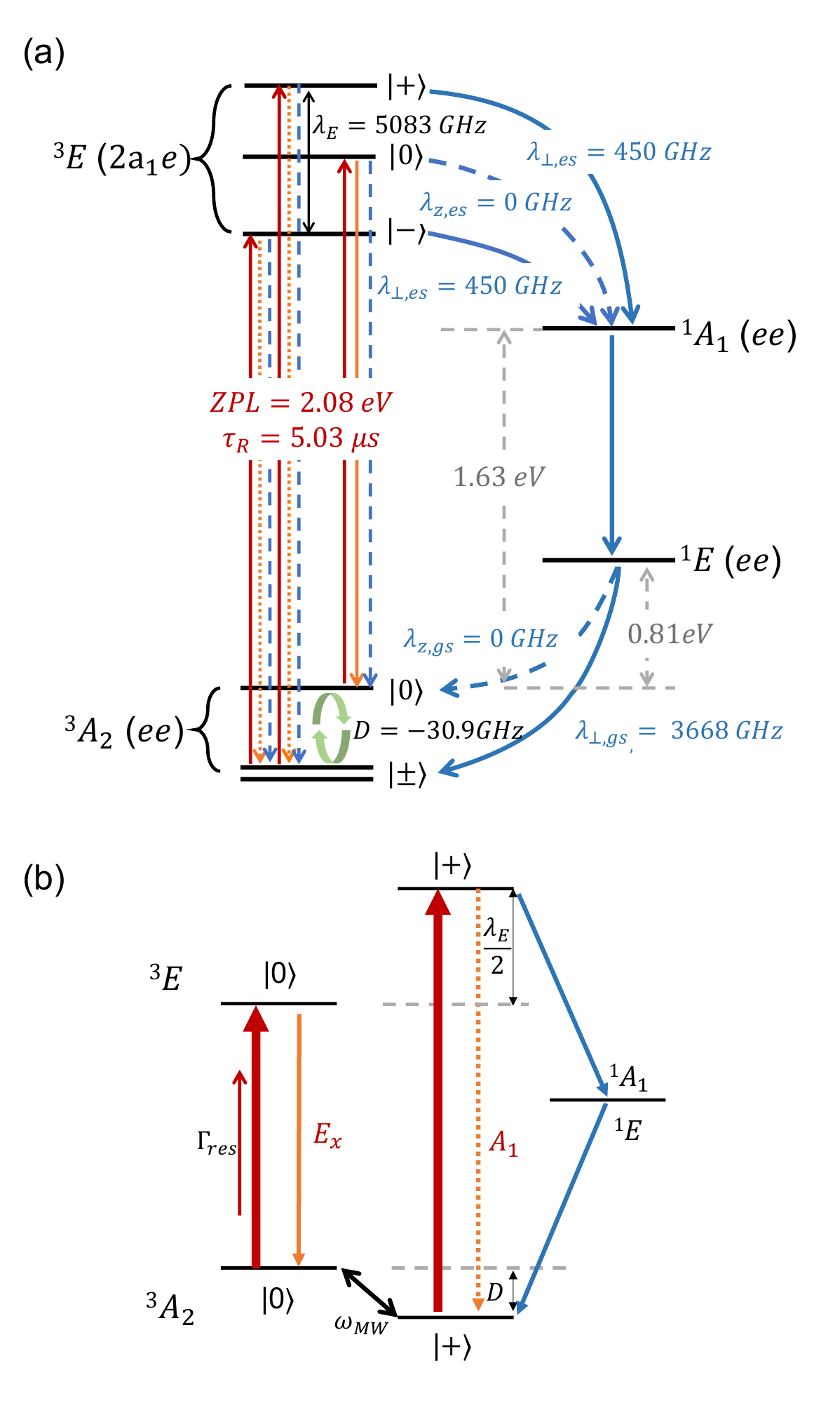}
    \caption{Spin properties and optical read-out of $\movo$ in ZnO. (a) Energy diagram and excited-state dynamics of $\movo$ in ZnO. (b) A schematic diagram for single shot readout with qubit encoded in $\ket{0}$ and $\ket{+}$. (For qubit encoded in $\ket{0}$ and $\ket{-}$, the only difference is that the $\lambda_E/2$ become negative.)
    $\Gamma_{\text{res}}$ denotes resonant excitation of the $\ket{0}$ state or the $\ket{\pm1}$ state. 
    In both panels (a) and (b), the red lines represent the optical excitation, the blue lines represent the phonon-assisted nonradiative recombination processes, and the orange lines represent the radiative recombination processes. Solid lines indicate dominant processes, dashed lines indicate forbidden ones, and dotted lines indicate allowed but less dominant processes. In panel (a), we show all these processes, while in panel (b), we keep only the allowed ones. }
    
    \label{fig9}
\end{figure}

Figure \ref{fig9}(a) provides the multiplet structure diagram of $\movo$ in ZnO that consists of a triplet ground state ($\gs$), triplet excited state ($\tes$), and two singlet shelving states ($\sgs$ and $\fses$), a 4-level structure similar to $\text{NV}^{-}$ in diamond~\cite{mazeGroupTheoreticalAnalysis2011}. 

The degeneracy of the ground state spin triplet manifold is split by intrinsic spin-spin and spin-orbit interactions. This splitting is characterized by the axial zero-field splitting (ZFS) parameter D (between $\ket{m_s=0}$ and $\ket{m_s=\pm1}$)  and  rhombic E (between $\ket{m_s=+1}$ and $\ket{m_s=-1}$). E is zero at $C_{3v}$ symmetry. We obtain the ground state ZFS by perturbation theory with first-order spin-spin contribution~\cite{smartIntersystemCrossingExciton2021,raysonFirstPrinciplesMethod2008b} $\text{D}_{SS} = 4.7$ GHz and second-order spin-orbit contribution\cite{neeseCalculationZerofieldSplitting2007}  $\text{D}_{SOC} = -35.6$ GHz, respectively. In  comparison, we obtain the $\text{D}_{SOC}+\text{D}_{SS}=21.00$ GHz for  $\vvo$, in reasonable agreement with the experimental value of 22.37 GHz~\cite{coffmanSpinHamiltonianParametersSpinOrbit1971,filipovichElectronParamagneticResonance1970} for $\text{V}^{3+}$ in ZnO (See Sec. X in the Supplemental Material~\cite{SI}). 

The spin-orbit coupling (SOC) parameters can be used to determine the fine structure of the spin state and the spin selectivity of optical transitions~\cite{ivadyFirstPrinciplesCalculation2018}. 
At the $C_{3v}$ symmetry, the spin-orbit coupling Hamiltonian can be written in terms of its axial($\lambda_z$) and nonaxial ($\lambda_\perp$) components: 
\begin{equation}
    H_{soc}=\frac{\lambda_{\perp}}{2}(L_+S_-+L_-S_+)+\lambda_zL_zS_z
\end{equation}
where $L_{\pm},L_z$ and $S_{\pm},S_z$ are the ladder operators of orbital and spin angular momentum, respectively, in the two-electron spin system with $\ket{S,m}$ basis. The axial component $\lambda_z$ can split the degenerate sub-levels within the excited-state triplet manifold~\cite{mazeGroupTheoreticalAnalysis2011}. It is also associated with the transitions between the triplet $m=0$ spin sub-levels and the singlet state. 
The non-axial component $\lambda_\perp$ is associated with the transitions between the triplet $m=\pm1$ spin sub-levels and the singlet state.

Given the multi-reference nature of excited states of $\movo$, we calculate the SOC matrix elements with the CASSCF method, which was shown to provide accurate SOC for multi-reference states~\cite{liExcitedstateDynamicsOptically2024a}. The computed ZFS and SOC parameters are shown in Table~\ref{tab_soc}.
In contrast to the ground state, where the first-order SOC contribution vanishes, the spin splitting of the $\tes$ excited state is dominated by a strong first-order SOC interaction, with $\lambda_E = \bra{^3E^{\pm}}H_{SOC}\ket{^3E^{\mp}} = 5083$ GHz, consistent with the group theory prediction~\cite{mazeGroupTheoreticalAnalysis2011}. We noted that the ratios  $\lambda_E (^3E)/D_{\mathrm{SOC}}(^3A_2)$ for both the NV center and $\movo$ are on the order of \( 10^2 \)~\cite{liExcitedstateDynamicsOptically2024a} in Table~\ref{tab_soc}.

\begin{table*}[!ht]
\centering
\begin{ruledtabular}
\caption{Calculated ZFS and SOC parameters of $\movo$ and NV center in diamond using SA(6)-CASSCF(6,6)~\cite{liExcitedstateDynamicsOptically2024a}. All values are in GHz. The $\lambda$ are the SOC parameters between states, where $\lambda_E = \bra{^3E^{\pm}}H_{SOC}\ket{^3E^{\mp}}$, $\lambda_{z,gs}=\bra{\sgs}H_{SOC}\ket{\gs^{0}}$, and  $\lambda_{\perp,gs}=\bra{\sgs}H_{SOC}\ket{\gs^{\pm}}$, $\lambda_{z,es}=\bra{^3E^{0}}H_{SOC}\ket{\fses}$, $\lambda_{\perp,es}=\bra{^3E^{\pm}}H_{SOC}\ket{\fses}$. The superscript on the right side of each state denotes the spin sublevel \( m_s \).}
\renewcommand{\arraystretch}{1.2}
\begin{tabular}{c|cccccccc}
              & $D_{SS}$ & $D_{SOC}$ &D & $\lambda_E$ & $\lambda_{z,gs}$ & $\lambda_{\perp,gs}$ & $\lambda_{z,es}$ & $\lambda_{\perp,es}$ \\    \hline
     $\movo$ &  4.70    & -35.60 & -30.90 & 5082.68 & 0.02 & 3668.05 & 0.02 & 450.48 \\ 
     NV$^-$~\cite{liExcitedstateDynamicsOptically2024a,smartIntersystemCrossingExciton2021} & 3.03 & 0.04 & 3.08 &14.21 & 0.03 & 5.22 & 0.06 & 3.96\\ 
\end{tabular}

\label{tab_soc}
\end{ruledtabular}
\end{table*}  

We find large non-axial component for both excited state and ground state intersystem crossings as shown in the Table~\ref{tab_soc} and Figure~\ref{fig9}. 
This large SOC originates from the transition metal ion Mo$^{4+}$ as a part of the vacancy defect.
The corresponding SOC parameters in NV center are two orders smaller than $\movo$. This indicates that the ISC in $\movo$ could be faster than the NV center in diamond, as the large SOC likely leads to a fast ISC~\cite{liExcitedstateDynamicsOptically2024a}.
The axial components $\lambda_{z,es}$ and $\lambda_{z,gs}$, on the other hand, are found to be zero, consistent with the group theory analysis of $C_{3v}$ systems~\cite{lenefElectronicStructureCenter1996,dohertyNegativelyChargedNitrogenvacancy2011}. This leads to the forbidden ISC transition to the spin $\ket{0}$ states.

The non-radiative ISC rates $\gamma$ between the initial (i) and final (j) states can be expressed by  first-order perturbation theory with the Fermi's Golden rule, i.e. $\gamma_{ij}=4\pi\hbar\lambda^2\tilde{X}_{ij}$, where $\lambda=\bra{i}H_{SOC}\ket{j}$ is the SOC coupling parameter between state $i$ and $j$, and $\tilde{X}_{ij}$ is the phonon coupling term in Eq.~\ref{eq:isc-F}~\cite{smartIntersystemCrossingExciton2021}. 
We further estimate the ISC rate for the spin $\ket{\pm}$ channel using the effective phonon approach
as detailed in Sec. XII of the Supplemental Material~\cite{SI}. 

The ISC lifetime (the inverse of the rate) corresponding to $\lambda_{\perp,gs}$ and $\lambda_{\perp,es}$ are found to be $\text{ISC}_{\perp,gs}=$0.25 ns and $\text{ISC}_{\perp,es}=$95.20 ps, respectively --- both are shorter than the spin-conserved radiative lifetime between triplet states. 

In contrast to NV center, the dynamic and pseudo Jahn–Teller (JT) effects are not observed in our system, i.e. the high-symmetry structure remains stable against symmetry-lowering distortion. We attribute this to the strong spin–orbit coupling (SOC) in the system, which quenches the Jahn–Teller effect and suppresses structural distortion~\cite{streltsovJahnTellerEffectSpinOrbit2020,bersukerJahnTellerEffect2006}. This quenching can be qualitatively understood using a spin-coupled "Mexican triplet hat" Jahn–Teller model, which is derived and discussed in detail in the SI~\cite{SI} section XIII.

\section{Discussion}

Based on the spin multiplet structure and excited-state dynamics of the $\movo$ defect in Figure~\ref{fig9}(a), we next discuss its optical read-out capability as a spin qubit.
While $\movo$ exhibits a similar four-level structure as the NV center in diamond, it offers unique advantages for high-fidelity single-shot spin readout and direct coherent control.  

These advantages arise from its significantly stronger spin–orbit coupling (SOC) compared to the NV center~\cite{razinkovasVibrationalVibronicStructure2021,bersukerJahnTellerPseudoJahnTeller2017,thieringTheoryOpticalSpinpolarization2018,thieringInitioMagnetoOpticalSpectrum2018}.

The schematic diagram for single shot readout is shown in the Figure~\ref{fig9}(b), where the $E_x$ and $A_1$ refer to transitions between the ground state(GS, $^3A_2$) and excited state(ES, $^3E$) within the spin channel $\ket{0}$ and $\ket{+}$, respectively. They consist of optical excitation(red), radiative transition(orange), and non-radiative recombination(blue). The spin-conserved non-radiative processes could be electron-phonon mediated processes as discussed in section D, or SOC-mediated ISC paths ($\tes^{\ket{\pm}}\rightarrow \fses\rightarrow \sgs\rightarrow \gs^{\ket{\pm}}$ for $A_1$ and $\tes^{\ket{0}}\rightarrow \fses\rightarrow \sgs\rightarrow \gs^{\ket{0}}$ for $E_x$). The cross lines represent the spin-flip process, where the spin-flip from $\ket{0}$ of $\tes$ to $\ket{\pm}$ of $\gs$($\gamma_{\perp,01}$) could arise from the ISC recombination path of $\tes^{\ket{0}}\rightarrow \fses\rightarrow \sgs\rightarrow \gs^{\ket{\pm}}$; similarly, the spin-flip from $\ket{\pm}$ ES to $\ket{0}$ GS ($\gamma_{\perp,10}$) could arise from the ISC recombination path of $\tes^{\ket{\pm}}\rightarrow \fses\rightarrow \sgs\rightarrow \gs^{\ket{0}}$. 

The preserved $C_{3v}$ symmetry in the absence of Jahn–Teller distortion leads to spin-selective ISC, and results in the following consequences:
(a) fast non-radiative recombination in $A_1$ (solid blue line) due to large $\lambda_{\perp,\text{es/gs}}$; (b) forbidden non-radiative recombination in $E_x$ (dashed blue line) due to $\lambda_{z,\text{gs/es}} = 0$; (c) forbidden spin-flip transitions: $\gamma_{\perp,01} = 0$ (caused by $\lambda_{z,\text{es}} = 0$) and $\gamma_{\perp,10} = 0$ (caused by $\lambda_{z,\text{gs}} = 0$).

First, due to $C_{3v}$ symmetry, the first-order SOC vanishes in $\gs$ but remains finite in the $\tes$. Combined with strong SOC, this result in a pronounced difference between the ground and excited triplet state spin splitting, which also leads to a large optical transition energy difference, i.e the energy difference between the $E_x$ and $A_1$ transitions in Figure~\ref{fig9}(b), enabling resonant excitation control even at high temperatures.
Furthermore, the spin-resolved photoluminescence (PL) peaks from these two transitions are inherently distinguishable by intensity: the $E_x$ related optical peak is expected to be significantly brighter than those of $A_1$. For the $E_x$, the fully forbidden non-radiative recombination leading to dominant spin-conserved radiative recombination ---hence, strong PL intensity. In contrast, the $A_1$ transitions include fast non-radiative ISC that competes with radiative recombination, resulting in dimmer PL peaks.

Finally, the forbidden spin-flip transitions of $\gamma_{\perp,01}$ and $\gamma_{\perp,10}$ lead to highly cyclic optical transitions that support high-fidelity single-shot spin readout. Moreover, the readout through the $E_x$ transition is especially robust at elevated temperatures and magnetic fields, because the large spin splitting of the excited state suppresses spin mixing between its sublevels, resulting in minimal $\gamma_{\perp,01}$. Specifically, resonant excitation of the $E_x$ allows for an effectively unlimited number of photon cycles, as $\gamma_{\perp,01}$ remains strictly forbidden for this state. The spin readout fidelity via resonant excitation of $E_x$ can be defined as $F_{00} = 1 - F_{\pm0}$, where $F_{\pm0}$ is the probability of obtaining the measurement outcome $\ket{\pm}$  state after optically pumping into $\ket{0}$ state~\cite{robledoHighfidelityProjectiveReadout2011}. The spin-flip probability $p = \frac{\gamma_{\perp,01}}{\gamma_{\perp,01} + \tau_R^{-1}}$, which constitutes the main source of error $F_{\pm0}$~\cite{christleIsolatedSpinQubits2017}, is also diminished, resulting in high spin readout fidelity. Such high fidelity single-shot readout enable the spin initialization of a single defect center. Once the spin state is read out, arbitrary spin initialization can be performed via radio-frequency (RF) control.

\section{Summary}
In conclusion, we have computationally identified $\movo$ as a deep-level defect qubit in ZnO with the essential ingredients for high-fidelity single-shot readout and long spin coherence time that are robust against environmental fluctuations. Our systematic screening based on thermodynamic stability and optical transition property calculations reveals that $\movo$ is an exceptional defect, which combines a spin-triplet ground state (S = 1), bright visible-range emission with high quantum yield, and an unusually small Huang–Rhys factor ($\sim$4.9), leading to a sharp zero-phonon line in ZnO. Spin decoherence analysis further reveals long $T_2$ times approaching 4 ms with a critical threshold paramagnetic impurity concentration of 0.035 ppm that marks the transition from nuclear- to electron-spin–dominated decoherence. 

We also examined the electronic spin multiplet structure and dynamics of $\movo$, including spin–orbit coupling, zero-field splitting, the Jahn–Teller effect, and intersystem crossing. Distinctively, this defect combines strong spin–orbit coupling with the absence of Jahn–Teller distortion, which enables high-fidelity single-shot readout and direct coherent spin control with minimal spin leakage at elevated temperatures. Together, these results establish $\movo$ as the first deep-level spin qubit candidate in ZnO, opening a pathway toward oxide-based quantum technologies that combine room-temperature operation, photonic integrability, and scalable defect engineering.

\section{Computational Methods}

\subsection{First-principles calculations}\label{First_principle_method}
We performed the ground state density-functional theory calculations using 
the Heyd–Scuseria–Ernzerhof (HSE) hybrid functional~\cite{heydHybridFunctionalsBased2003}, by the plane-wave code Vienna Ab initio Simulation Package (VASP)~\cite{kresseEfficientIterativeSchemes1996,kresseEfficiencyAbinitioTotal1996,kresseInitioMoleculardynamicsSimulation1994}.  The multideterminant constrained DFT (mcDFT) were used for geometry relaxations of the excited states.
The defect was constructed and relaxed in a 4x4x3 supercell of ZnO. We use plane-wave cutoff of 400 eV and the projector-augmented wave (PAW) poseudopotentials to conduct the structural relaxations.  
We choose the Fock exchange fraction parameter of 0.375 and screening parameter $\omega=$0.2 $\textup{~\AA}$, which reproduces the experimental lattice constants and band gap of ZnO~\cite{thomasSemiconductorsBasicData1997,AlbertssonZnOExperiment, ObaDefectEnergetics, ReynoldsZnOExperimentbandgap}. 
We obtain a band gap of 3.43 eV, and a lattice parameter (a=3.249$\textup{~\AA}$, c=5.204$\textup{~\AA}$) of the pristine zinc oxide, in good agreement with experimental data (band gap 3.4 eV, lattice constants a=3.25$\textup{~\AA}$ and c=5.21$\textup{~\AA}$).

We use the r2SCAN functional for phonon calculations, which is known for its reliability in capturing the static and dynamical properties of the lattice with a low computational cost~\cite{ningReliableLatticeDynamics2022}. We use the Hubbard U correction of 2 eV for the 3d orbital of Zn and obtain the phonon band with very good agreement with experiments (see Figure~S10~\cite{SI}).

\subsection{Defect Formation Energy and Charge Transition Level}
The charged defect formation energy $E_f^q(d)$ of the defect $d$ with charge state $q$ is calculated by
\begin{equation}
    E_f^q(d)=E^q_{tot}(d)-E_{tot}(p)-\sum_i\Delta N_i\mu_i+qE_F+E_{corr}
\label{defect_FE}
\end{equation}
where $E^q_{tot}(d)$ is the total energy of the supercell containing the defect with charge state $q$, and $E_{tot}(p)$ is the total energy of the pristine system in the same supercell as the defect. $\Delta N_i$ denotes the difference in the number of atoms of type $i$ between the defect and pristine systems ($\Delta N_i>0$ means an atom of type $i$ has been added to the defect system and $\Delta N_i<0$ means that an atom has been removed). $\mu_i$ and $E_F$ are the chemical potential of species $i$ and the Fermi energy, respectively. $E_{corr}$ is charged cell correction to eliminate the fictitious Coulomb interaction of defects with its own periodic images and fictitious homogeneous compensating background~\cite{sundararamanFirstprinciplesElectrostaticPotentials2017,wuFirstprinciplesEngineeringCharged2017}.

In $\xvo$ defect complex, the third term in Eq. \ref{defect_FE} can be written as $\sum_i\Delta N_i\mu_i=-\mu_{\text{Zn}}-\mu_{\text{O}}+\mu_{\text{X}}$. The formation energy of each defect is calculated with chemical potentials in two conditions: O-rich and O-poor. In the O-rich condition the oxygen chemical potential is computed by total energy of the oxygen molecule ($\mu_\text{O}^{\text{O-rich}}=1/2E_{tot}(\text{O}_2)$). In the O-poor condition the zinc chemical potential is computed by $\mu_{\text{Zn}}^{\text{O-poor}}=E_{tot}(\text{Zn})$ where $E_{tot}(\text{Zn})$ is the energy of zinc crystal~\cite{janottiNativePointDefects2007}. The $\mu_\text{O}^{\text{O-poor}}$ and $\mu_{\text{Zn}}^{\text{O-rich}}$ is computed from $\mu_\text{Zn}^{\text{O-poor}}$ and $\mu_{\text{O}}^{\text{O-rich}}$ according to the constraint $\mu_\text{Zn}^{\text{O-poor/rich}}+\mu_\text{O}^{\text{O-poor/rich}}=\mu_{\text{ZnO}}$.
The chemical potential of the dopant X (X = Ti, Nb, V, Mo) is computed from its most stable oxide compound, where $\mu_{\text{Ti}}$, $\mu_{\text{Nb}}$, $\mu_{\text{V}}$, and $\mu_{\text{Mo}}$ are computed from $\text{TiO}_2$, $\text{Nb}_2\text{O}_5$, $\text{V}_2\text{O}_3$, and $\text{Mo}\text{O}_2$, respectively~\cite{obaDesignExplorationSemiconductors2018}. 

The thermodynamic charge transition level (CTL) between charge states $q$ and $q'$ ($\epsilon(q/q')$) is the Fermi-level position at which the formation energy of charge state $q$ and $q'$ are the same~\cite{obaDesignExplorationSemiconductors2018,wuFirstprinciplesEngineeringCharged2017}:
\begin{equation}
    \varepsilon(q/q')=\frac{E_f^q(E_F=0)-E_f^{q'}(E_F=0)}{q'-q}
\end{equation}
Here, $E_f^q(E_F=0)$ is the defect formation energy of the defect at the Fermi-level $E_F=0$ (aligned with VBM). The defect system is more stable at a charge state $q$ when its Fermi-level is smaller than CTL $\varepsilon(q/q')$ and is more stable at q' when $E_F>\varepsilon(q/q')$.

\subsection{Absorption spectrum and radiative lifetime}
The optical absorption spectrum at the random-phase approximation (RPA) is computed with the Yambo code\cite{onidaElectronicExcitationsDensityfunctional2002},  including the local field effect for the polarizability, with the input single particle states from hybrid functional HSE (Fock exchange = 0.375)~\cite{sangalliManybodyPerturbationTheory2019}.  The DFT single particle states are calculated by the open source plan-wave code Quantum Espresso~\cite{giannozziQUANTUMESPRESSOModular2009} with norm-conserving Vanderbilt (ONCV) pseudopotentials~\cite{hamannOptimizedNormconservingVanderbilt2013,schlipfOptimizationAlgorithmGeneration2015} and wavefunction cutoff of 80Ry. We then extract the transition dipole moment and oscillator strength for each excitation. 
The radiative lifetime for defects transition is derived from the Fermi's golden rule~\cite{wuDimensionalityAnisotropicityDependence2019b,smartIntersystemCrossingExciton2021} and computed by:

\begin{equation}\label{equation_rad}
    \tau_R=\frac{3\pi\epsilon_0 h^4 c^3}{n_D e^2E^3\mu^2}
\end{equation}
where $E$ is the excitation energy, $c$ is the speed of light, $\mu^2$ is the modulus square of transition dipole moment, and $n_D=\sqrt{\epsilon}=2.4$ is reflective index computed from the dielectric constant of pristine ZnO.

\subsection{Multideterminant cDFT\label{sec-mcDFT}}
The idea of multideterminant cDFT is that the energy expectation value of a linear combination of states is a linear combination of the energies of the pure states. This idea has been employed on describing the multi-reference nature for spin defects in past works~\cite{czelejQuantumBehaviorHydrogenvacancy2018, czelejTransitionMetalRelatedQuantumEmitters2024, galiInitioSupercellCalculations2008, shangFirstprinciplesStudyTransition2022,ivanovElectronicExcitationsCharged2023}. 
For the $\movo$ and $\nbvo$, the two-particle wavefunction constructed from the two highest occupied $\ket{e_{x/y}}$ orbitals can be written as: 
\begin{subequations}
\begin{eqnarray}
        &\ket{^3A_2^{m_s=1}}=\frac{1}{\sqrt{2}}(\ket{e_xe_y}-\ket{e_ye_x})\\
        &\ket{^3A_2^{m_s=0}}=\frac{1}{\sqrt{2}}(\ket{e_x\bar{e}_y}-\ket{e_y\bar{e}_x})\\
        &\ket{^3A_2^{m_s=-1}}=\frac{1}{\sqrt{2}}(\ket{\bar{e}_x\bar{e}_y}-\ket{\bar{e}_y\bar{e}_x})     \\
        &\ket{^1E_x}=\frac{1}{\sqrt{2}}(\ket{e_x\bar{e}_y}+\ket{e_y\bar{e}_x})\\
        &\ket{^1E_y}=\frac{1}{\sqrt{2}}(\ket{e_x\bar{e}_x}-\ket{e_y\bar{e}_y})
\end{eqnarray}
\end{subequations}
where the "$\bar{e}_{x/y}$" represent the occupied spin minority $e_{x/y}$ state, and the "$e_{x/y}$" represent the occupied spin majority $e_{x/y}$. The triplet ground states $\ket{^3A_2^{m_s=\pm1}}$ and $\ket{^3A_2^{m_s=0}}$ should have degenerate energy $E(\ket{^3A_2})$, we therefore use the single configuration at $m_s=1$ to compute the ground state energy: $E(\ket{^3A_2})=E(\ket{e_xe_y})$. The degenerated triplet excited state $^3E_{x/y}$ (correspond to the spin majority $e\rightarrow  a_1$ transition) can also be computed from the single determinant: $E(\ket{^3E_{x/y}})=E(\ket{2a_1e_{x/y}})$. In the calculation, we use the fractional occupation $E(\ket{2a_1e})=E(\ket{2a_1e_x^{0.5}e_y^{0.5}})$ to preserve the state degeneracy. The $E(\ket{^3A_2'}) = E(\ket{\bar{a}_1 e_x e_y 2\bar{a}_1})$ state corresponds to an $a_1 \rightarrow 2a_1$ excitation and is essentially single-determinantal in character, allowing it to be treated directly within a standard cDFT framework.

The DFT close shell calculation for the $\movo$ broke the degeneracy of the $e_{x/y}$ orbitals, resulting in the spin-symmetry-broken determinant~\cite{ivanovElectronicExcitationsCharged2023}: $\ket{e_-\bar{e}_+}$ and the corresponded energy $E(e_-\bar{e}_+)$ where $e_+=\frac{e_x+e_y}{\sqrt{2}}$ and $e_-=\frac{e_x-e_y}{\sqrt{2}}$. 

Notice $\ket{e_-\bar{e}_+}=\frac{1}{2}\ket{(e_x-e_y)(\bar{e}_x+\bar{e}_y)}=\frac{1}{\sqrt{2}}(\ket{^1E_y}+\ket{^3A_2^{m_s=0}})$. We therefore have: 
\begin{subequations}
\begin{eqnarray}
    &E(e_-\bar{e}_+)=\bra{e_-\bar{e}_+}H\ket{e_-\bar{e}_+}\\
    &=\frac{1}{2}(E(^1E_y)+E(^3A_2^{m_s=0}))\nonumber \\
    &\Rightarrow E(\ket{^1E})=2E(\ket{e_-\bar{e}_+})-E(\ket{e_xe_y})
\end{eqnarray}
\end{subequations}

\subsection{Non-radiative lifetime, Photoluminescence and Huang Rhy's factor}\label{sec-Non-rad}

The nonradiative lifetime ($\tau^{NR}$) is a measure of how fast the nonradiative recombination happens between the final state $\ket{f}$ and initial state $\ket{i}$. The phonon-assisted nonradiative recombination is evaluated via Fermi's golden rule as below, 
\begin{align}
    \frac{1}{\tau^{NR}_{if}} &= \frac{2\pi}{\hbar}g\sum_{n,m}p_{in}|\mel{fm}{H^{e-ph}}{in}|^2\delta(E_{fm}-E_{in}) 
    \label{eq:NR}
\end{align}
where $H^{e-ph}$ is the electron-phonon coupling Hamiltonian, $g$ is the degeneracy factor of the final state that depends on the number of equivalent atomic configurations, and $p_{in}$ is the occupation number of the vibronic state $\ket{in}$ following the Boltzmann distribution.

Under the static coupling approximation with one-dimensional (1D) phonon approximation~\cite{smartIntersystemCrossingExciton2021, wuCarrierRecombinationMechanism2019}, we can rewrite Eq.~(\ref{eq:NR}) as
\begin{align}
\frac{1}{\tau^{NR}_{if}}
    &= \frac{2\pi}{\hbar}g|W_{if}|^2X_{if}(T) \label{eq:nonrad_wx}\\
    W_{if} &= \mel{\psi_i(\mathbf{r, R})}{\frac{\partial H}{\partial Q}}{\psi_f(\mathbf{r, R})}|_{\mathbf{R=R_a}} \label{eq:wif}\\
    X_{if} &= \sum_{n,m}p_{in}|\mel{\phi_{fm}(\mathbf{R})}{Q-Q_a}{\phi_{in}(\mathbf{R})}|^2 \nonumber\\
    &\times\delta(m\hbar\omega_f-n\hbar\omega_i+\Delta E_{if}). \label{eq:xif}
\end{align}
Eq.~(\ref{eq:nonrad_wx}) is separated into the electronic term ($W_{if}$), which depends on the electronic wave function ($\psi$) overlap, and the phonon term ($X_{if}$), which describes the strength of the phonon contribution. $W_{if}$ is determined using finite differences of Kohn-Sham orbitals from DFT calculation using HSE(0.375) functional.
The phonon term $X_{if}$ includes the energy conservation between the initial and final vibronic states with vibrational frequencies of $\omega_i$ and $\omega_f$, and $\phi$ is the phonon wave function obtained from harmonic oscillator wavefunctions. The detailed derivation can be found in Refs.~\cite{alkauskasFirstprinciplesTheoryNonradiative2014b, wuCarrierRecombinationMechanism2019}.

To validate the 1D phonon approximation, we compare the Huang-Rhys factor calculated with the 1D effective phonon and full phonon results, as detailed in the TABLE S8~\cite{SI}.
To compute the intersystem crossing (ISC) rate, we adopted the approach derived from nonradiative rates, as implemented in our in-house code~\cite{smartIntersystemCrossingExciton2021,liExcitedstateDynamicsOptically2024a}:
\begin{equation}
    \Gamma_{ISC} = 4 \pi \hbar \lambda_\perp^2 \widetilde{X}_{if}(T) \label{eq:isc-full}
\end{equation}
\begin{equation}
\begin{split}
    \widetilde{X}_{if}(T) = \sum_{n,m}p_{in}  &\left|
        \braket{\phi_{fm}(\textbf{R})}{\phi_{in}(\mathbf{R})}
    \right|^2 \times \\
    & \delta(m\hbar\omega_f - n\hbar\omega_i+\Delta E_{if}).
\end{split}
\label{eq:isc-F}
\end{equation}

This method allows different values for the initial-state vibrational frequency ($\omega_i$) and the final-state frequency ($\omega_f$) through explicit calculations of the phonon wavefunction overlap, and we obtain good agreement with experimental values of ISC rates for NV center in diamond~\cite{smartIntersystemCrossingExciton2021, liExcitedstateDynamicsOptically2024a}.

The photoluminescence lineshape spectrum was simulated using the Huang-Rhys method with all phonon eigenmodes, implemented in our in-house codes~\cite{liCarbonTrimerEV2022, alkauskasFirstPrinciplesCalculationsLuminescence2012a}, with $\gamma=0.005 $eV and smearing = 0.003 eV, where $\gamma$ is a free parameter that accounts for the broadening of PL. Due to the well-known failure of local and semilocal exchange correlation functional on the ZnO system~\cite{mccluskeyChapterEightPoint2015, obaPointDefectsZnO2011} and the high computational cost of the hybrid functional, we apply the r2SCAN functional with Hubbard U for the all phonon calculations using the Phonopy code interfaced with VASP and Hubbard U corrections of $U_{Zn-d}= 2 eV$ and $U_{Nb-d}= 3.7 eV$. It has been tested that this method reproduces the defect local structure comparable to HSE as well as the phonon band of pristine ZnO (see Figure~S10~\cite{SI}).

\subsection{Decoherence time}
\subsection*{Quantum bath model to compute the spin decoherence}

We employ the quantum bath theory to compute the spin decoherence \cite{seoQuantumDecoherenceDynamics2016, parkDecoherenceNitrogenvacancySpin2022, leeFirstprinciplesTheoryExtending2022a}, in which the decoherence occurs due to the entanglement between a central spin and its environment. We consider electronic and nuclear spin baths as the environment. Bath spins are randomly distributed in the lattice and bath spins, within a certain radius from the defect qubit, are included in the calculation. This bath radius ($r_\text{bath}$) is determined by performing a systematic convergence test as shown in Figure~S12~\cite{SI}. We find that a radius of 5 nm gives a numerically converged result for the nuclear spin bath. For the electronic spin bath, we find that the bath radii of 16, 37, 85, and 220 nm are appropriate for 100, 10, 1, and 0.1 ppm of electronic spin concentrations, respectively. 

In addition, we use another parameter ($r_\text{dipole}$), which sets the maximum distance for the interaction between bath spins. This means that if two bath spins are separated by a distance larger than $r_\text{dipole}$, the two spins are considered as non-interacting. We find that our CCE calculations are converged with $r_\text{dipole}$ of 1 nm for the nuclear spin bath, and for the electronic spin bath, $r_\text{dipole}$ of 13, 33, 62, and 120 nm for 100, 10, 1, and 0.1 ppm of electronic spin concentrations.

The dynamics of the total system of qubit and environment is governed by a spin Hamiltonian, which is expressed as:

\begin{equation}
    H = H_d + H_\text{bath} + H_{d-\text{bath}}
\end{equation}

where $H_d$ is the qubit Hamiltonian, $H_\text{bath}$ is the bath spin Hamiltonian, and $H_{d-\text{bath}}$ is the interaction between the qubit and the bath spins. For the nuclear spin bath, each component of the Hamiltonian is given as follows:

\begin{equation}
\begin{aligned}
    H_d &= -\gamma_e \mathbf{B} \cdot \mathbf{S},  \\
    H_\text{bath} &= -\mathbf{B} \cdot \sum_i \gamma_{n_i} \mathbf{I}_i + H_{n-n} + H_Q, \\
    H_{d-\text{bath}} &= S_z \sum_i \mathbf{A}_i \cdot \mathbf{I}_i \\
    &= \sum_i \left( B_{ix} I_{ix} S_z + B_{iy} I_{iy} S_z + A_{i} I_{iz} S_z \right), \\
    H_{n-n} &= \frac{\mu_0}{4\pi} \sum_{i,j} \gamma_{n_i} \gamma_{n_j} \left[ 
    \frac{\mathbf{I}_i \cdot \mathbf{I}_j}{r_{ij}^3} - 
    \frac{3 (\mathbf{I}_i \cdot \mathbf{r}_{ij})(\mathbf{I}_j \cdot \mathbf{r}_{ij})}{r_{ij}^5} \right],  \\
    H_Q &= \frac{e Q}{6I(2I-1)} \sum_{\alpha,\beta} V_{\alpha\beta} 
    \left[ \frac{3}{2} \left( I_\alpha I_\beta + I_\beta I_\alpha \right) - \delta_{\alpha\beta} \mathbf{I}^2 \right]. 
\end{aligned}
\end{equation}

In these equations, $\gamma_e$ and $\gamma_{n_i}$ are the gyromagnetic ratios of the electron spin ($\mathbf{S}$) and the $i$-th nuclear spin ($\mathbf{I}_i$) in the bath, respectively. The external magnetic field is aligned parallel to the defect's symmetry axis. We adopt the secular approximation for the hyperfine interaction ($H_{d-\text{bath}}$), in which the non-secular terms including $S_x$ and $S_y$ are neglected. In the nuclear spin-spin interaction ($H_{n-n}$), $\mu_0$ is the vacuum permeability and $r_{ij}$ is the distance between the $i$-th nuclear spin and the $j$-th nuclear spin.

$H_Q$ is the nuclear quadrupole interaction, in which $eQ$ is the quadrupole moment of the isotope under consideration interacting with the EFG tensor $V_{\alpha\beta}$ ($\alpha, \beta = x, y, z$).

The hyperfine tensor ($\mathbf{A}$) and the electric field gradient (EFG) tensor ($V_{\alpha\beta}$) are computed by using DFT as implemented in Vienna Ab initio Simulation Package (VASP) code at HSE hybrid functional level of theory. 

The spin Hamiltonian for the electronic spin bath can be expressed as follows:

\begin{equation}
\begin{aligned}
    H_d &= -\gamma_e \mathbf{B} \cdot \mathbf{S},  \\
    H_\text{bath} &= -\mathbf{B} \sum_i \gamma_{e_i} \mathbf{S}_i + H_{e-e},  \\
    H_{e-e} &= -\frac{\mu_0}{4\pi} \sum_{i,j} \gamma_{e_i} \gamma_{e_j} \left[ 
    \frac{\mathbf{S}_i \cdot \mathbf{S}_j}{r_{ij}^3} - 
    \frac{3 (\mathbf{S}_i \cdot \mathbf{r}_{ij})(\mathbf{S}_j \cdot \mathbf{r}_{ij})}{r_{ij}^5} \right], 
\end{aligned}
\end{equation}

Similar to the case of the nuclear spin bath, the qubit Hamiltonian $H_d$ represents the Zeeman interaction of the qubit (electron spin) with an external magnetic field $\mathbf{B}$. Here, $\gamma_e$ is the gyromagnetic ratio of the electron and $S_z$ is the $z$-component of the electron spin operator. The bath spin Hamiltonian describes the dynamics of the bath spins (other electron spins in the environment). The first term of the bath spin Hamiltonian represents the Zeeman interaction of each electron spin in the bath with the external magnetic field. And $H_{e-e}$ in the bath spin Hamiltonian represents the magnetic dipolar interactions between the electron spins in the bath. Here, $\mathbf{S}_i$ represents the spin operator of the $i$-th electronic spin in the bath.

The coherence function $\mathcal{L}(t)$ is given as the off-diagonal element of the reduced density matrix, formally expressed as:

\begin{equation}
    \mathcal{L}(t) \equiv \frac{\mathrm{tr}[\rho_\text{tot}(t) S_+]}{\mathrm{tr}[\rho_\text{tot}(0) S_+]}
\end{equation}

where $\rho_\text{tot}(t)$ is the total density matrix of the qubit ($\rho_e$) and the bath ($\rho_\text{bath}$) at time $t$. $S_+$ is the electron spin raising operator, defined as $S_+ = S_x + i S_y$. To compute the coherence function, we employ the cluster correlation expansion (CCE) technique \cite{yangQuantumManybodyTheory2008a}, which enables a systematic expansion of the coherence function in many-body systems. We find that second-order CCE (CCE-2) and first-order CCE (CCE-1) give numerically converged results for the Hahn-echo decay time ($T_2$) and the FID time ($T_2^*$), respectively.

\subsection{ZFS and SOC strength}

The zero-field splitting (ZFS) parameters (D and E) consist of first-order spin-spin interaction contributions ($\text{D}_{SS}$,$\text{E}_{SS}$) and second-order spin-orbit (SO) contributions ($\text{D}_{SOC}$, $\text{E}_{SOC}$ )~\cite{neeseCalculationZerofieldSplitting2007}. For transition metals, the SO contribution is significant and cannot be neglected. We computed the SO contribution to ZFS using the linear response method implemented in ORCA~\cite{neeseORCAProgramSystem2012,neeseORCAQuantumChemistry2020}, employing all-electron PBE0 calculations with the def2-TZVP basis set~\cite{weigendBalancedBasisSets2005}. In contrast, the spin-spin contribution to ZFS, which is relatively long-ranged compared to the SO contribution, did not converge with the cluster size (see TABLE S10~\cite{SI}). To address this, we calculated the spin-spin contribution using our in-house code~\cite{smartIntersystemCrossingExciton2021} interfaced with the plane-wave code QUANTUM ESPRESSO~\cite{giannozziQUANTUMESPRESSOModular2009}.

We use the complete-active-space self-consistent field (CASSCF) method implemented in the ORCA code~\cite{neeseORCAProgramSystem2012,neeseORCAQuantumChemistry2020} to obtain SOC matrix elements between the electronic states. The second order Douglas-Kroll-Hess (DKH2) Hamiltonian~\cite{reiher2006dkh}, the all-electron DKH-def2-TZVP basis set ~\cite{pantazis2008all}, and the SARC-DKH-TZVP basis set~\cite{rolfes2020all} are used to account for the scalar relativistic effects. The spin-orbit mean-field operator is used for the SOC calculation~\cite{neese2005soc}. The cluster is created by retaining atoms near the defects and passivate the surface dangling bond with pseudo hydrogen atoms with core charge of q=(8-m)/4~\cite{huangSurfacePassivationMethod2005}
, where we choose the O-H bond to be 1.057 $\textup{~\AA}$ and Zn-H to be 1.731 $\textup{~\AA}$~\cite{badaevaInvestigationPureCo22008}.
The perturbative correction with N-electron
Valence State Perturbation Theory (NEVPT2)~\cite{guoSparseMapsSystematicInfrastructure2016} was applied upon the CASSCF to obtain the multi-reference excited state with both static and dynamic correlation. More details can be found in SI section XI~\cite{SI}.

\section*{Data availability}

The input files for the simulations, Python post-processing scripts, example output files, and source data used in this study are available in Ref.~\cite{ZnOdataBase}.

\section*{Code availability}

The codes for electronic structure are available through open-source software QUANTUM ESPRESSO\citep{giannozziQUANTUMESPRESSOModular2009}, the WEST-code\citep{govoniLargeScaleGW2015}, and the ORCA code. The in-house open source codes for calculating nonradiative and intersystem crossing rates\cite{wuCarrierRecombinationMechanism2019,smartIntersystemCrossingExciton2021,pingComputationalDesignQuantum2021} are available from github \url{https://github.com/Ping-Group-UCSC/NonRad-ISC}.

\section*{Author contributions}
S. Zhang, E. Perez, K. Li performed first-principles calculations of ground and excited state properties, and dynamical properties for spin-photon interfaces. Y. Wang, J. Bazantes, R. Zhang, J. Sun provided inputs for SCAN calculations. T. Park and H. Seo performed the spin coherence time calculations. X. Wang, and K. Mei provided inputs for results' interpretation and experimental applications. S. Zhang, E. Perez, T. Park, K. Fu, H. Seo and Y. Ping wrote the first draft of this paper. All authors contributed to the writing of the manuscript. Y.Ping designed and supervised all aspects of the study.

\section*{Competing interests}

The authors declare no competing interests.

\section*{Additional information}

\textbf{Supplementary information} is available for this paper at {[}url{]}.

\textbf{Correspondence} and requests for materials should be addressed to K.F., H.S. and Y.P.

\begin{acknowledgments}
We acknowledge the very helpful discussion with Andrei Faraon, Joe Falson, and Juan Carlos Idrobo. We acknowledge the support by AFOSR CFIRE program under grant FA9550-23-1-0418. Ping also acknowledges the support by the National Science Foundation under grant no. DMR-2143233 for the support of computational technique development in this work.
This research used resources of the Scientific Data and Computing center, a component of the 
Computational Science Initiative, at Brookhaven National Laboratory under Contract No. DE-SC0012704,
the National Energy Research Scientific Computing Center (NERSC) a U.S. Department of Energy Office of Science User Facility operated under Contract No. DE-AC02-05CH11231. 
This work used the TACC Stampede3 system at the University of Texas at Austin through allocation PHY240212 from the Advanced Cyberinfrastructure Coordination Ecosystem: Services and Support (ACCESS) program~\cite{Boerner2023}, which is supported by US National Science Foundation grants No. 2138259, No. 2138286, No. 2138307, No. 2137603, and No. 2138296.
TP and HS were supported by the National Research Foundation (NRF) of Korea grant funded by the Korean government (MSIT) (No. 2023R1A2C1006270, No. RS-2025-25454922), by Creation of the Quantum Information Science R\&D Ecosystem (No. RS-2023-NR068116) through the NRF, by the KIST institutional program (No. 2E33571), by Institute of Information \& communications Technology Planning \& Evaluation (IITP) grant funded by the MSIT (RS-2025-25464252), and by the education and training program of the Quantum Information Research Support Center funded through the NRF of Korea (No.2021M3H3A103657313).
YW, JDVB, RZ, and JS acknowledge the support by the U.S. DOE, Office of Science, Basic Energy Sciences (BES), Grant No. DE-SC0014208.
\end{acknowledgments}

\bibliographystyle{apsrev4-2}
\bibliography{ref}

@article{Tsukazaki_2008,
doi = {10.1143/APEX.1.055004},
url = {https://doi.org/10.1143/APEX.1.055004},
year = {2008},
month = {may},
publisher = {},
volume = {1},
number = {5},
pages = {055004},
author = {Tsukazaki, Atsushi and Yuji, Hiroyuki and Akasaka, Shunsuke and Tamura, Kentaro and Nakahara, Ken and Tanabe, Tetsuhiro and Takasu, Hidemi and Ohtomo, Akira and Kawasaki, Masashi},
title = {High Electron Mobility Exceeding 104 cm2 V-1 s-1 in MgxZn1-xO/ZnO Single Heterostructures Grown by Molecular Beam Epitaxy},
journal = {Applied Physics Express}
}

@misc{ZnOdataBase,
  title        = {Input files, post-processing scripts, and source data for this paper are available in https://github.com/Ping-Group-UCSC/ZnO\_project\_data },
  year         = {2026},
  url = {https://github.com/Ping-Group-UCSC/ZnO_project_data}
}

@misc{SI,
  note = "See Supplemental Material at
    URL-will-be-inserted-by-publisher for additional discussion, analysis, and numerical data."
}

@ARTICLE{Sukachev2017-rr,
  title     = "Silicon-vacancy spin qubit in Diamond: A quantum memory exceeding
               10 ms with single-shot state readout",
  author    = "Sukachev, D D and Sipahigil, A and Nguyen, C T and Bhaskar, M K
               and Evans, R E and Jelezko, F and Lukin, M D",
  journal   = "Phys. Rev. Lett.",
  publisher = "American Physical Society (APS)",
  volume    =  119,
  number    =  22,
  month     =  nov,
  year      =  2017,
  language  = "en"
}

@ARTICLE{Koehl2011-iv,
  title     = "Room temperature coherent control of defect spin qubits in
               silicon carbide",
  author    = "Koehl, William F and Buckley, Bob B and Heremans, F Joseph and
               Calusine, Greg and Awschalom, David D",
  journal   = "Nature",
  publisher = "Springer Science and Business Media LLC",
  volume    =  479,
  number    =  7371,
  pages     = "84--87",
  month     =  nov,
  year      =  2011,
  language  = "en"
}

@inproceedings{Boerner2023,
author = {Boerner, Timothy J. and Deems, Stephen and Furlani, Thomas R. and Knuth, Shelley L. and Towns, John},
title = {ACCESS: Advancing Innovation: NSF’s Advanced Cyberinfrastructure Coordination Ecosystem: Services \& Support},
year = {2023},
isbn = {9781450399852},
publisher = {Association for Computing Machinery},
address = {New York, NY, USA},
url = {https://doi.org/10.1145/3569951.3597559},
doi = {10.1145/3569951.3597559},
booktitle = {Practice and Experience in Advanced Research Computing 2023: Computing for the Common Good},
pages = {173–176},
numpages = {4},
location = {Portland, OR, USA},
series = {PEARC '23}
}

@article{pingComputationalDesignQuantum2021,
  title = {Computational Design of Quantum Defects in Two-Dimensional Materials},
  author = {Ping, Yuan and Smart, Tyler J.},
  year = {2021},
  month = oct,
  journal = {Nat. Comput. Sci.},
  volume = {1},
  number = {10},
  pages = {646--654},
  publisher = {Nature Publishing Group},
  issn = {2662-8457},
  doi = {10.1038/s43588-021-00140-w},
  urldate = {2022-03-01},
  copyright = {2021 Springer Nature America, Inc.},
  langid = {english},
}

@article{awschalomQuantumTechnologiesOptically2018,
  title = {Quantum Technologies with Optically Interfaced Solid-State Spins},
  author = {Awschalom, David D. and Hanson, Ronald and Wrachtrup, J{\"o}rg and Zhou, Brian B.},
  year = {2018},
  month = sep,
  journal = {Nat. Photon},
  volume = {12},
  number = {9},
  pages = {516--527},
  publisher = {Nature Publishing Group},
  issn = {1749-4893},
  doi = {10.1038/s41566-018-0232-2},
  urldate = {2024-01-03},
  copyright = {2018 Springer Nature Limited},
  langid = {english},

}

@article{weberQuantumComputingDefects2010,
  title = {Quantum Computing with Defects},
  author = {Weber, J. R. and Koehl, W. F. and Varley, J. B. and Janotti, A. and Buckley, B. B. and Van De Walle, C. G. and Awschalom, D. D.},
  year = {2010},
  month = may,
  journal = {Proc. Natl. Acad. Sci. U.S.A.},
  volume = {107},
  number = {19},
  pages = {8513--8518},
  issn = {0027-8424, 1091-6490},
  doi = {10.1073/pnas.1003052107},
  urldate = {2024-12-19},
}

@article{castellettoSiliconCarbideColor2020,
  title = {Silicon Carbide Color Centers for Quantum Applications},
  author = {Castelletto, Stefania and Boretti, Alberto},
  year = {2020},
  month = mar,
  journal = {J. Phys. Photonics},
  volume = {2},
  number = {2},
  pages = {022001},
  publisher = {IOP Publishing},
  issn = {2515-7647},
  doi = {10.1088/2515-7647/ab77a2},
  urldate = {2024-11-10},
  langid = {english},
}

@article{gongCoherentDynamicsStrongly2023,
  title = {Coherent Dynamics of Strongly Interacting Electronic Spin Defects in Hexagonal Boron Nitride},
  author = {Gong, Ruotian and He, Guanghui and Gao, Xingyu and Ju, Peng and Liu, Zhongyuan and Ye, Bingtian and Henriksen, Erik A. and Li, Tongcang and Zu, Chong},
  year = {2023},
  month = jun,
  journal = {Nat. Commun},
  volume = {14},
  number = {1},
  pages = {3299},
  publisher = {Nature Publishing Group},
  issn = {2041-1723},
  doi = {10.1038/s41467-023-39115-y},
  urldate = {2025-02-01},
  copyright = {2023 The Author(s)},
  langid = {english}
}

@article{gottschollInitializationReadoutIntrinsic2020a,
  title = {Initialization and Read-out of Intrinsic Spin Defects in a van Der {{Waals}} Crystal at Room Temperature},
  author = {Gottscholl, Andreas and Kianinia, Mehran and Soltamov, Victor and Orlinskii, Sergei and Mamin, Georgy and Bradac, Carlo and Kasper, Christian and Krambrock, Klaus and Sperlich, Andreas and Toth, Milos and Aharonovich, Igor and Dyakonov, Vladimir},
  year = {2020},
  month = may,
  journal = {Nat. Mater.},
  volume = {19},
  number = {5},
  pages = {540--545},
  publisher = {Nature Publishing Group},
  issn = {1476-4660},
  doi = {10.1038/s41563-020-0619-6},
  urldate = {2025-02-01},
  copyright = {2020 The Author(s), under exclusive licence to Springer Nature Limited},
  langid = {english}

}

@article{gottschollRoomTemperatureCoherent2021a,
  title = {Room Temperature Coherent Control of Spin Defects in Hexagonal Boron Nitride},
  author = {Gottscholl, Andreas and Diez, Matthias and Soltamov, Victor and Kasper, Christian and Sperlich, Andreas and Kianinia, Mehran and Bradac, Carlo and Aharonovich, Igor and Dyakonov, Vladimir},
  year = {2021},
  month = apr,
  journal = {Sci. Adv},
  volume = {7},
  number = {14},
  pages = {eabf3630},
  publisher = {American Association for the Advancement of Science},
  doi = {10.1126/sciadv.abf3630},
  urldate = {2025-02-01},

}

@article{harrisCoherenceGroupIVColor2024,
  title = {Coherence of Group-{{IV}} Color Centers},
  author = {Harris, Isaac B. W. and Englund, Dirk},
  year = {2024},
  month = feb,
  journal = {Phys. Rev. B},
  volume = {109},
  number = {8},
  pages = {085414},
  publisher = {American Physical Society},
  doi = {10.1103/PhysRevB.109.085414},
  urldate = {2025-02-01},
}

@article{karapatzakisMicrowaveControlTinVacancy2024,
  title = {Microwave {{Control}} of the {{Tin-Vacancy Spin Qubit}} in {{Diamond}} with a {{Superconducting Waveguide}}},
  author = {Karapatzakis, Ioannis and Resch, Jeremias and Schrodin, Marcel and Fuchs, Philipp and Kieschnick, Michael and Heupel, Julia and Kussi, Luis and S{\"u}rgers, Christoph and Popov, Cyril and Meijer, Jan and Becher, Christoph and Wernsdorfer, Wolfgang and Hunger, David},
  year = {2024},
  month = aug,
  journal = {Phys. Rev. X},
  volume = {14},
  number = {3},
  pages = {031036},
  publisher = {American Physical Society},
  doi = {10.1103/PhysRevX.14.031036},
  urldate = {2025-02-01},
}

@article{orphal-kobinCoherentMicrowaveOptical,
  title = {Coherent {{Microwave}}, {{Optical}}, and {{Mechanical Quantum Control}} of {{Spin Qubits}} in {{Diamond}}},
  author = {{Orphal-Kobin}, Laura and Torun, Cem G{\"u}ney and Bopp, Julian M. and Pieplow, Gregor and Schr{\"o}der, Tim},
  journal = {Adv. Quantum Technol.},
  volume = {n/a},
  number = {n/a},
  pages = {2300432},
  issn = {2511-9044},
  doi = {10.1002/qute.202300432},
  urldate = {2025-02-01},
  copyright = {{\copyright} 2024 The Authors. Advanced Quantum Technologies published by Wiley-VCH GmbH},
  langid = {english},
  year = {2024}
}

@article{tarasenkoSpinOpticalProperties2018,
  title = {Spin and {{Optical Properties}} of {{Silicon Vacancies}} in {{Silicon Carbide}} - {{A Review}}},
  author = {Tarasenko, S. A. and Poshakinskiy, A. V. and Simin, D. and Soltamov, V. A. and Mokhov, E. N. and Baranov, P. G. and Dyakonov, V. and Astakhov, G. V.},
  year = {2018},
  journal = {Phys. Status Solidi B},
  volume = {255},
  number = {1},
  pages = {1700258},
  issn = {1521-3951},
  doi = {10.1002/pssb.201700258},
  urldate = {2022-06-16},
  langid = {english},

}

@article{andersonFivesecondCoherenceSingle2022,
  title = {Five-Second Coherence of a Single Spin with Single-Shot Readout in Silicon Carbide},
  author = {Anderson, Christopher P. and Glen, Elena O. and Zeledon, Cyrus and Bourassa, Alexandre and Jin, Yu and Zhu, Yizhi and Vorwerk, Christian and Crook, Alexander L. and Abe, Hiroshi and {Ul-Hassan}, Jawad and Ohshima, Takeshi and Son, Nguyen T. and Galli, Giulia and Awschalom, David D.},
  year = {2022},
  month = feb,
  journal = {Science Advances},
  volume = {8},
  number = {5},
  pages = {eabm5912},
  publisher = {American Association for the Advancement of Science},
  doi = {10.1126/sciadv.abm5912},
  urldate = {2025-03-31}
}

@article{christleIsolatedSpinQubits2017,
  title = {Isolated {{Spin Qubits}} in {{SiC}} with a {{High-Fidelity Infrared Spin-to-Photon Interface}}},
  author = {Christle, David J. and Klimov, Paul V. and De Las Casas, Charles F. and Sz{\'a}sz, Kriszti{\'a}n and Iv{\'a}dy, Viktor and Jokubavicius, Valdas and Ul Hassan, Jawad and Syv{\"a}j{\"a}rvi, Mikael and Koehl, William F. and Ohshima, Takeshi and Son, Nguyen T. and Janz{\'e}n, Erik and Gali, {\'A}d{\'a}m and Awschalom, David D.},
  year = {2017},
  month = jun,
  journal = {Phys. Rev. X},
  volume = {7},
  number = {2},
  pages = {021046},
  issn = {2160-3308},
  doi = {10.1103/PhysRevX.7.021046},
  urldate = {2025-03-31},
  copyright = {https://creativecommons.org/licenses/by/4.0/},
  langid = {english}
}

@article{hopperSpinReadoutTechniques2018,
  title = {Spin {{Readout Techniques}} of the {{Nitrogen-Vacancy Center}} in {{Diamond}}},
  author = {Hopper, David and Shulevitz, Henry and Bassett, Lee},
  year = {2018},
  month = aug,
  journal = {Micromachines},
  volume = {9},
  number = {9},
  pages = {437},
  issn = {2072-666X},
  doi = {10.3390/mi9090437},
  urldate = {2025-07-02},
  langid = {english},

}

@article{irberRobustAllopticalSingleshot2021,
  title = {Robust All-Optical Single-Shot Readout of Nitrogen-Vacancy Centers in Diamond},
  author = {Irber, Dominik M. and Poggiali, Francesco and Kong, Fei and Kieschnick, Michael and L{\"u}hmann, Tobias and Kwiatkowski, Damian and Meijer, Jan and Du, Jiangfeng and Shi, Fazhan and Reinhard, Friedemann},
  year = {2021},
  month = jan,
  journal = {Nat Commun},
  volume = {12},
  number = {1},
  pages = {532},
  publisher = {Nature Publishing Group},
  issn = {2041-1723},
  doi = {10.1038/s41467-020-20755-3},
  urldate = {2025-07-02},
  copyright = {2021 The Author(s)},
  langid = {english}
}

@article{robledoHighfidelityProjectiveReadout2011,
  title = {High-Fidelity Projective Read-out of a Solid-State Spin Quantum Register},
  author = {Robledo, Lucio and Childress, Lilian and Bernien, Hannes and Hensen, Bas and Alkemade, Paul F. A. and Hanson, Ronald},
  year = {2011},
  month = sep,
  journal = {Nature},
  volume = {477},
  number = {7366},
  pages = {574--578},
  publisher = {Nature Publishing Group},
  issn = {1476-4687},
  doi = {10.1038/nature10401},
  urldate = {2025-07-03},
  copyright = {2011 Springer Nature Limited},
  langid = {english},
}

@article{rosenthalSingleShotReadoutWeak2024,
  title = {Single-{{Shot Readout}} and {{Weak Measurement}} of a {{Tin-Vacancy Qubit}} in {{Diamond}}},
  author = {Rosenthal, Eric I. and Biswas, Souvik and Scuri, Giovanni and Lee, Hope and Stein, Abigail J. and Kleidermacher, Hannah C. and Grzesik, Jakob and Rugar, Alison E. and Aghaeimeibodi, Shahriar and Riedel, Daniel and Titze, Michael and Bielejec, Edward S. and Choi, Joonhee and Anderson, Christopher P. and Vu{\v c}kovi{\'c}, Jelena},
  year = {2024},
  month = oct,
  journal = {Phys. Rev. X},
  volume = {14},
  number = {4},
  pages = {041008},
  issn = {2160-3308},
  doi = {10.1103/PhysRevX.14.041008},
  urldate = {2025-03-28},
  langid = {english}
}

@article{kanaiGeneralizedScalingSpin2022a,
  title = {Generalized Scaling of Spin Qubit Coherence in over 12,000 Host Materials},
  author = {Kanai, Shun and Heremans, F. Joseph and Seo, Hosung and Wolfowicz, Gary and Anderson, Christopher P. and Sullivan, Sean E. and Onizhuk, Mykyta and Galli, Giulia and Awschalom, David D. and Ohno, Hideo},
  year = {2022},
  month = apr,
  journal = {Proc. Natl. Acad. Sci. U.S.A.},
  volume = {119},
  number = {15},
  pages = {e2121808119},
  issn = {0027-8424, 1091-6490},
  doi = {10.1073/pnas.2121808119},
  urldate = {2024-12-19},
  langid = {english},
}

@article{falsonMagnesiumDopingControlled2011,
  title = {Magnesium {{Doping Controlled Density}} and {{Mobility}} of {{Two-Dimensional Electron Gas}} in {{MgxZn1-xO}}/{{ZnO Heterostructures}}},
  author = {Falson, Joseph and Maryenko, Denis and Kozuka, Yusuke and Tsukazaki, Atsushi and Kawasaki, Masashi},
  year = {2011},
  month = aug,
  journal = {Appl. Phys. Express},
  volume = {4},
  number = {9},
  pages = {091101},
  publisher = {IOP Publishing},
  issn = {1882-0786},
  doi = {10.1143/APEX.4.091101},
  urldate = {2024-12-19},
  langid = {english}
}

@article{falsonMgZnOZnOHeterostructures2016,
  title = {{{MgZnO}}/{{ZnO}} Heterostructures with Electron Mobility Exceeding 1\,{\texttimes}\,106\,Cm2/{{Vs}}},
  author = {Falson, Joseph and Kozuka, Yusuke and Uchida, Masaki and Smet, Jurgen H. and Arima, Taka-hisa and Tsukazaki, Atsushi and Kawasaki, Masashi},
  year = {2016},
  month = may,
  journal = {Scientific Reports},
  volume = {6},
  number = {1},
  pages = {26598},
  issn = {2045-2322},
  doi = {10.1038/srep26598},
}

@article{liElectronMobilityZnMgO2013,
  title = {Electron {{Mobility}} in {{ZnMgO}}/{{ZnO Heterostructures}} in the {{Bloch}}--{{Gr{\"u}neisen Regime}}},
  author = {Li, Qun and Zhang, Jingwen and Chong, Jing and Hou, Xun},
  year = {2013},
  month = dec,
  journal = {Appl. Phys. Express},
  volume = {6},
  number = {12},
  pages = {121102},
  publisher = {IOP Publishing},
  issn = {1882-0786},
  doi = {10.7567/APEX.6.121102},
  urldate = {2024-12-19},
  langid = {english}
}

@book{haynes2010crc,
  title={CRC Handbook of Chemistry and Physics, 91st Edition},
  author={Haynes, W.M.},
  isbn={9781439820735},
  url={https://books.google.com/books?id=w8stkgEACAAJ},
  year={2010},
  publisher={Taylor \& Francis Group},
  address={Boca Raton, FL}
}

@article{shannonRevisedEffectiveIonic1976,
  title = {Revised Effective Ionic Radii and Systematic Studies of Interatomic Distances in Halides and Chalcogenides},
  author = {Shannon, R. D.},
  year = {1976},
  month = sep,
  journal = {Acta Cryst A},
  volume = {32},
  number = {5},
  pages = {751--767},
  publisher = {International Union of Crystallography},
  issn = {0567-7394},
  doi = {10.1107/S0567739476001551},
  urldate = {2024-05-24},
}

@article{seoDesigningDefectbasedQubit2017,
  title = {Designing Defect-Based Qubit Candidates in Wide-Gap Binary Semiconductors for Solid-State Quantum Technologies},
  author = {Seo, Hosung and Ma, He and Govoni, Marco and Galli, Giulia},
  year = {2017},
  month = dec,
  journal = {Phys. Rev. Mater.},
  volume = {1},
  number = {7},
  pages = {075002},
  issn = {2475-9953},
  doi = {10.1103/PhysRevMaterials.1.075002},
  urldate = {2024-05-24},
  copyright = {https://link.aps.org/licenses/aps-default-license},
  langid = {english}
}

@article{niaourisEnsembleSpinRelaxation2021,
  title = {Ensemble spin relaxation of shallow donor qubits in ZnO},
  author = {Niaouris, Vasileios and Durnev, Mikhail V. and Linpeng, Xiayu and Viitaniemi, Maria L. K. and Zimmermann, Christian and Vishnuradhan, Aswin and Kozuka, Yusuke and Kawasaki, Masashi and Fu, Kai-Mei C.},
  journal = {Phys. Rev. B},
  volume = {105},
  issue = {19},
  pages = {195202},
  numpages = {9},
  year = {2022},
  month = {May},
  publisher = {American Physical Society},
  doi = {10.1103/PhysRevB.105.195202},
  url = {https://link.aps.org/doi/10.1103/PhysRevB.105.195202}
}

@article{NiaourisLinewidth2024,
author = {Vasileios Niaouris and Samuel H. D'Ambrosia and Christian Zimmermann and Xingyi Wang and Ethan R. Hansen and Michael Titze and Edward S. Bielejec and Kai-Mei C. Fu},
journal = {Opt. Quantum},
number = {1},
pages = {7--13},
publisher = {Optica Publishing Group},
title = {Contributions to the optical linewidth of shallow donor-bound excitonic transition in ZnO},
volume = {2},
month = {Feb},
year = {2024},
url = {https://opg.optica.org/opticaq/abstract.cfm?URI=opticaq-2-1-7},
doi = {10.1364/OPTICAQ.501568}}

@article{viitaniemiCoherentSpinPreparation2022,
  title = {Coherent {{Spin Preparation}} of {{Indium Donor Qubits}} in {{Single ZnO Nanowires}}},
  author = {Viitaniemi, Maria L. K. and Zimmermann, Christian and Niaouris, Vasileios and D’Ambrosia, Samuel H. and Wang, Xingyi and Kumar, E. Senthil and Mohammadbeigi, Faezeh and Watkins, Simon P. and Fu, Kai-Mei C.},
  date = {2022-03-09},
  journal = {Nano Lett.},
  volume = {22},
  number = {5},
  pages = {2134--2139},
  publisher = {American Chemical Society},
  issn = {1530-6984},
  doi = {10.1021/acs.nanolett.1c04156},
  url = {https://doi.org/10.1021/acs.nanolett.1c04156},
  urldate = {2024-05-24},
}

@article{wangPropertiesDonorQubits2023,
  title = {Properties of {{Donor Qubits}} in {{Zn O Formed}} by {{Indium-Ion Implantation}}},
  author = {Wang, Xingyi and Zimmermann, Christian and Titze, Michael and Niaouris, Vasileios and Hansen, Ethan R. and D’Ambrosia, Samuel H. and Vines, Lasse and Bielejec, Edward S. and Fu, Kai-Mei C.},
  date = {2023-05-30},
  journal = {Phys. Rev. Applied},
  volume = {19},
  number = {5},
  pages = {054090},
  issn = {2331-7019},
  doi = {10.1103/PhysRevApplied.19.054090},
  url = {https://link.aps.org/doi/10.1103/PhysRevApplied.19.054090},
  urldate = {2024-05-24},
  langid = {english},
}

@article{linpengCoherencePropertiesShallow2018,
  title = {Coherence {{Properties}} of {{Shallow Donor Qubits}} in {{Zn O}}},
  author = {Linpeng, Xiayu and Viitaniemi, Maria L.K. and Vishnuradhan, Aswin and Kozuka, Y. and Johnson, Cameron and Kawasaki, M. and Fu, Kai-Mei C.},
  year = {2018},
  month = dec,
  journal = {Phys. Rev. Applied},
  volume = {10},
  number = {6},
  pages = {064061},
  issn = {2331-7019},
  doi = {10.1103/PhysRevApplied.10.064061},
  urldate = {2022-02-02},
  langid = {english},
}

@article{dohertyNitrogenvacancyColourCentre2013,
  title = {The Nitrogen-Vacancy Colour Centre in Diamond},
  author = {Doherty, Marcus W. and Manson, Neil B. and Delaney, Paul and Jelezko, Fedor and Wrachtrup, J{\"o}rg and Hollenberg, Lloyd C. L.},
  year = {2013},
  month = {Jul},
  journal = {Phys. Rep.},
  volume = {528},
  number = {1},
  pages = {1--45},
  issn = {0370-1573},
  doi = {10.1016/j.physrep.2013.02.001},
  urldate = {2023-08-04},
  langid = {english},
}

@article{mazeGroupTheoreticalAnalysis2011,
  title = {Group Theoretical Analysis of Nitrogen-Vacancy Center's Energy Levels and Selection Rules},
  author = {Maze, J. R. and Gali, A. and Togan, E. and Chu, Y. and Trifonov, A. and Kaxiras, E. and Lukin, M. D.},
  year = {2011},
  journal = {MRS Proc.},
  volume = {1282},
  pages = {mrsf10-1282-a07-03},
  issn = {0272-9172, 1946-4274},
  doi = {10.1557/opl.2011.310},
  urldate = {2024-10-30},
  copyright = {https://www.cambridge.org/core/terms},
  langid = {english},

}

@article{grayLocalAtomicEnvironment2017,
  title = {Local Atomic Environment of the {{Cu-related}} Defect in Zinc Oxide},
  author = {Gray, Ciar{\'a}n and Trefflich, Lukas and R{\"o}der, Robert and Ronning, Carsten and Henry, Martin O and McGlynn, Enda},
  year = {2017},
  month = apr,
  journal = {J. Phys. D: Appl. Phys.},
  volume = {50},
  number = {14},
  pages = {145105},
  issn = {0022-3727, 1361-6463},
  doi = {10.1088/1361-6463/aa5f07},
  urldate = {2025-03-05},
}

@article{haoStructuralOpticalMagnetic2012,
  title = {Structural, Optical, and Magnetic Studies of Manganese-Doped Zinc Oxide Hierarchical Microspheres by Self-Assembly of Nanoparticles},
  author = {Hao, Yao-Ming and Lou, Shi-Yun and Zhou, Shao-Min and Yuan, Rui-Jian and Zhu, Gong-Yu and Li, Ning},
  year = {2012},
  month = dec,
  journal = {Nanoscale Res Lett},
  volume = {7},
  number = {1},
  pages = {100},
  issn = {1556-276X},
  doi = {10.1186/1556-276X-7-100},
  urldate = {2025-03-05},
}

@article{liaoDcThermalPlasma2006,
  title = {Dc Thermal Plasma Synthesis and Properties of Zinc Oxide Nanorods},
  author = {Liao, Shih-Chieh and Lin, Hsiu-Fen and Hung, Sung-Wei and Hu, Chen-Ti},
  year = {2006},
  month = may,
  journal = {JVST B},
  volume = {24},
  number = {3},
  pages = {1322--1326},
  issn = {1071-1023},
  doi = {10.1116/1.2197513},
  urldate = {2025-03-05},
}

@article{shahroosvandSolutionbasedSyntheticStrategies2013,
  title = {Solution-Based Synthetic Strategies for {{Eu}} Doped {{ZnO}} Nanoparticle with Enhanced Red Photoluminescence},
  author = {Shahroosvand, Hashem and {Ghorbani-asl}, Mahsa},
  year = {2013},
  month = dec,
  journal = {J. Lumin.},
  volume = {144},
  pages = {223--229},
  issn = {0022-2313},
  doi = {10.1016/j.jlumin.2013.06.003},
  urldate = {2025-03-05},

}

@article{singhInvestigationLowtemperatureExcitonic2010,
  title = {Investigation of Low-Temperature Excitonic and Defect Emission from {{Ni-doped ZnO}} Nanoneedles and {{V-doped ZnO}} Nanostructured Film},
  author = {Singh, Shubra and Nakamura, Daisuke and Sakai, Kentaro and Okada, Tatsuo and Ramachandra Rao, M S},
  year = {2010},
  month = feb,
  journal = {New J. Phys.},
  volume = {12},
  number = {2},
  pages = {023007},
  issn = {1367-2630},
  doi = {10.1088/1367-2630/12/2/023007},
  urldate = {2025-03-05},

}

@incollection{mccluskeyChapterEightPoint2015,
  title = {Chapter {{Eight}} - {{Point Defects}} in {{ZnO}}},
  booktitle = {Semiconductors and {{Semimetals}}},
  author = {McCluskey, Matthew D.},
  editor = {Romano, Lucia and Privitera, Vittorio and Jagadish, Chennupati},
  year = {2015},
  month = jan,
  series = {Defects in {{Semiconductors}}},
  volume = {91},
  pages = {279--313},
  publisher = {Elsevier},
  doi = {10.1016/bs.semsem.2014.11.002},
  urldate = {2024-04-10},
}

@article{obaPointDefectsZnO2011,
  title = {Point Defects in {{ZnO}}: An Approach from First Principles},
  shorttitle = {Point Defects in {{ZnO}}},
  author = {Oba, Fumiyasu and Choi, Minseok and Togo, Atsushi and Tanaka, Isao},
  year = {2011},
  month = jun,
  journal = {STAM},
  volume = {12},
  number = {3},
  pages = {034302},
  publisher = {Taylor \& Francis},
  issn = {1468-6996},
  doi = {10.1088/1468-6996/12/3/034302},
  urldate = {2024-05-17},
  pmid = {27877390},
}

@article{mccluskeyDefectsZnO2009,
  title = {Defects in {{ZnO}}},
  author = {McCluskey, M. D. and Jokela, S. J.},
  year = {2009},
  month = oct,
  journal = {J. Appl. Phys.},
  volume = {106},
  number = {7},
  pages = {071101},
  issn = {0021-8979},
  doi = {10.1063/1.3216464},
  urldate = {2025-10-02}
}

@article{LishuVOntype,
  title = {Oxygen vacancies: The origin of $n$-type conductivity in ZnO},
  author = {Liu, Lishu and Mei, Zengxia and Tang, Aihua and Azarov, Alexander and Kuznetsov, Andrej and Xue, Qi-Kun and Du, Xiaolong},
  journal = {Phys. Rev. B},
  volume = {93},
  issue = {23},
  pages = {235305},
  numpages = {6},
  year = {2016},
  month = {Jun},
  publisher = {American Physical Society},
  doi = {10.1103/PhysRevB.93.235305}
}

@article{frodasonZnVacPolaron,
  title = {Zn vacancy as a polaronic hole trap in ZnO},
  author = {Frodason, Y. K. and Johansen, K. M. and Bj\o{}rheim, T. S. and Svensson, B. G. and Alkauskas, A.},
  journal = {Phys. Rev. B},
  volume = {95},
  issue = {9},
  pages = {094105},
  numpages = {8},
  year = {2017},
  month = {Mar},
  publisher = {American Physical Society},
  doi = {10.1103/PhysRevB.95.094105},
  url = {https://link.aps.org/doi/10.1103/PhysRevB.95.094105}
}

@article{clarkIntrinsicDefectsZnOHydrid,
  title = {Intrinsic defects in ZnO calculated by screened exchange and hybrid density functionals},
  author = {Clark, S. J. and Robertson, J. and Lany, S. and Zunger, A.},
  journal = {Phys. Rev. B},
  volume = {81},
  issue = {11},
  pages = {115311},
  numpages = {5},
  year = {2010},
  month = {Mar},
  publisher = {American Physical Society},
  doi = {10.1103/PhysRevB.81.115311},
  url = {https://link.aps.org/doi/10.1103/PhysRevB.81.115311}
}

@article{lyonsFirstprinciplesCharacterizationNativedefectrelated2017a,
  title = {First-Principles Characterization of Native-Defect-Related Optical Transitions in {{ZnO}}},
  author = {Lyons, J. L. and Varley, J. B. and Steiauf, D. and Janotti, A. and {Van de Walle}, C. G.},
  year = {2017},
  month = jul,
  journal = {J. Appl. Phys},
  volume = {122},
  number = {3},
  pages = {035704},
  issn = {0021-8979},
  doi = {10.1063/1.4992128},
  urldate = {2024-10-16},
}

@article{sundararamanFirstprinciplesElectrostaticPotentials2017,
  title = {First-Principles Electrostatic Potentials for Reliable Alignment at Interfaces and Defects},
  author = {Sundararaman, Ravishankar and Ping, Yuan},
  year = {2017},
  month = mar,
  journal = {J. Chem. Phys},
  volume = {146},
  number = {10},
  pages = {104109},
  issn = {0021-9606, 1089-7690},
  doi = {10.1063/1.4978238},
  urldate = {2021-06-07},
  langid = {english},
}

@article{wuFirstprinciplesEngineeringCharged2017,
  title = {First-Principles Engineering of Charged Defects for Two-Dimensional Quantum Technologies},
  author = {Wu, Feng and Galatas, Andrew and Sundararaman, Ravishankar and Rocca, Dario and Ping, Yuan},
  year = {2017},
  month = dec,
  journal = {Phys. Rev. Mater.},
  volume = {1},
  number = {7},
  pages = {071001},
  issn = {2475-9953},
  doi = {10.1103/PhysRevMaterials.1.071001},
  urldate = {2024-05-29},
  copyright = {https://link.aps.org/licenses/aps-default-license},
  langid = {english},
}

@article{obaDesignExplorationSemiconductors2018,
  title = {Design and Exploration of Semiconductors from First Principles: {{A}} Review of Recent Advances},
  shorttitle = {Design and Exploration of Semiconductors from First Principles},
  author = {Oba, Fumiyasu and Kumagai, Yu},
  year = {2018},
  month = jun,
  journal = {Appl. Phys. Express},
  volume = {11},
  number = {6},
  pages = {060101},
  issn = {1882-0778, 1882-0786},
  doi = {10.7567/APEX.11.060101},
  urldate = {2024-05-30},
  langid = {english},
}

@article{janottiNativePointDefects2007,
  title = {Native Point Defects in {{ZnO}}},
  author = {Janotti, Anderson and Van De Walle, Chris G.},
  year = {2007},
  month = oct,
  journal = {Phys. Rev. B},
  volume = {76},
  number = {16},
  pages = {165202},
  issn = {1098-0121, 1550-235X},
  doi = {10.1103/PhysRevB.76.165202},
  urldate = {2024-01-12},
  langid = {english},
}

@article{smartIntersystemCrossingExciton2021,
  title = {Intersystem Crossing and Exciton--Defect Coupling of Spin Defects in Hexagonal Boron Nitride},
  author = {Smart, Tyler J. and Li, Kejun and Xu, Junqing and Ping, Yuan},
  year = {2021},
  month = apr,
  journal = {npj Comput Mater},
  volume = {7},
  number = {1},
  pages = {1--8},
  publisher = {Nature Publishing Group},
  issn = {2057-3960},
  doi = {10.1038/s41524-021-00525-5},
  urldate = {2021-10-17},
  copyright = {2021 The Author(s)},
  langid = {english},
}

@article{wuDimensionalityAnisotropicityDependence2019b,
  title = {Dimensionality and Anisotropicity Dependence of Radiative Recombination in Nanostructured Phosphorene},
  author = {Wu, Feng and Rocca, Dario and Ping, Yuan},
  year = {2019},
  month = oct,
  journal = {J. Mater. Chem. C},
  volume = {7},
  number = {41},
  pages = {12891--12897},
  publisher = {The Royal Society of Chemistry},
  issn = {2050-7534},
  doi = {10.1039/C9TC02214G},
  urldate = {2022-07-11},
  langid = {english},

}

@article{maExcitedStatesNegatively2010a,
  title = {Excited States of the Negatively Charged Nitrogen-Vacancy Color Center in Diamond},
  author = {Ma, Yuchen and Rohlfing, Michael and Gali, Adam},
  year = {2010},
  month = jan,
  journal = {Phys. Rev. B},
  volume = {81},
  number = {4},
  pages = {041204},
  issn = {1098-0121, 1550-235X},
  doi = {10.1103/PhysRevB.81.041204},
  urldate = {2024-07-09},
  copyright = {http://link.aps.org/licenses/aps-default-license},
  langid = {english},
}

@article{onidaElectronicExcitationsDensityfunctional2002,
  title = {Electronic Excitations: Density-Functional versus Many-Body {{Green}}'s-Function Approaches},
  shorttitle = {Electronic Excitations},
  author = {Onida, Giovanni and Reining, Lucia and Rubio, Angel},
  year = {2002},
  month = jun,
  journal = {Rev. Mod. Phys.},
  volume = {74},
  number = {2},
  pages = {601--659},
  issn = {0034-6861, 1539-0756},
  doi = {10.1103/RevModPhys.74.601},
  urldate = {2021-11-05},
  langid = {english},
}

@article{wuCarrierRecombinationMechanism2019,
  title = {Carrier Recombination Mechanism at Defects in Wide Band Gap Two-Dimensional Materials from First Principles},
  author = {Wu, Feng and Smart, Tyler J. and Xu, Junqing and Ping, Yuan},
  year = {2019},
  month = aug,
  journal = {Phys. Rev. B},
  volume = {100},
  number = {8},
  pages = {081407},
  issn = {2469-9950, 2469-9969},
  doi = {10.1103/PhysRevB.100.081407},
  urldate = {2021-06-07},
  langid = {english},
}

@article{markhamInteractionNormalModes1959,
  title = {Interaction of {{Normal Modes}} with {{Electron Traps}}},
  author = {Markham, Jordan J.},
  year = {1959},
  month = oct,
  journal = {Rev. Mod. Phys.},
  volume = {31},
  number = {4},
  pages = {956--989},
  issn = {0034-6861},
  doi = {10.1103/RevModPhys.31.956},
  urldate = {2024-12-20},
  copyright = {http://link.aps.org/licenses/aps-default-license},
  langid = {english},
}

@article{walkerOpticalAbsorptionLuminescence1979,
  title = {Optical Absorption and Luminescence in Diamond},
  author = {Walker, J},
  year = {1979},
  month = oct,
  journal = {Rep. Prog. Phys.},
  volume = {42},
  number = {10},
  pages = {1605--1659},
  issn = {0034-4885, 1361-6633},
  doi = {10.1088/0034-4885/42/10/001},
  urldate = {2024-12-20},
  langid = {english},
}

@article{ningReliableLatticeDynamics2022,
  title = {Reliable {{Lattice Dynamics}} from an {{Efficient Density Functional Approximation}}},
  author = {Ning, Jinliang and Furness, James W. and Sun, Jianwei},
  year = {2022},
  month = mar,
  journal = {Chem. Mater.},
  volume = {34},
  number = {6},
  pages = {2562--2568},
  issn = {0897-4756, 1520-5002},
  doi = {10.1021/acs.chemmater.1c03222},
  urldate = {2024-10-20},
  copyright = {https://creativecommons.org/licenses/by/4.0/},
  langid = {english},
}

@article{alkauskasFirstprinciplesTheoryLuminescence2014,
  title = {First-Principles Theory of the Luminescence Lineshape for the Triplet Transition in Diamond {{NV}} Centres},
  author = {Alkauskas, Audrius and Buckley, Bob B. and Awschalom, David D. and de Walle, Chris G. Van},
  year = {2014},
  month = jul,
  journal = {New J. Phys.},
  volume = {16},
  number = {7},
  pages = {073026},
  publisher = {IOP Publishing},
  issn = {1367-2630},
  doi = {10.1088/1367-2630/16/7/073026},
  urldate = {2024-05-17},
}

@article{bulancea-lindvallChlorineVacancySiC2023,
  title = {Chlorine Vacancy in 4 {{H}} - {{SiC}} : {{An NV-like}} Defect with Telecom-Wavelength Emission},
  shorttitle = {Chlorine Vacancy in 4 {{H}} - {{SiC}}},
  author = {{Bulancea-Lindvall}, Oscar and Davidsson, Joel and Armiento, Rickard and Abrikosov, Igor A.},
  year = {2023},
  month = dec,
  journal = {Phys. Rev. B},
  volume = {108},
  number = {22},
  pages = {224106},
  issn = {2469-9950, 2469-9969},
  doi = {10.1103/PhysRevB.108.224106},
  urldate = {2024-10-20},
  langid = {english},

}

@article{ivadyInitioTheoryNegatively2020,
  title = {Ab Initio Theory of the Negatively Charged Boron Vacancy Qubit in Hexagonal Boron Nitride},
  author = {Iv{\'a}dy, Viktor and Barcza, Gergely and Thiering, Gerg{\H o} and Li, Song and Hamdi, Hanen and Chou, Jyh-Pin and Legeza, {\"O}rs and Gali, Adam},
  year = {2020},
  month = apr,
  journal = {npj Comput Mater},
  volume = {6},
  number = {1},
  pages = {1--6},
  publisher = {Nature Publishing Group},
  issn = {2057-3960},
  doi = {10.1038/s41524-020-0305-x},
  urldate = {2022-10-03},
}

@article{reimersPhotoluminescencePhotophysicsPhotochemistry2020,
  title = {Photoluminescence, Photophysics, and Photochemistry of the {{V B}} - Defect in Hexagonal Boron Nitride},
  author = {Reimers, Jeffrey R. and Shen, Jun and Kianinia, Mehran and Bradac, Carlo and Aharonovich, Igor and Ford, Michael J. and Piecuch, Piotr},
  year = {2020},
  month = oct,
  journal = {Phys. Rev. B},
  volume = {102},
  number = {14},
  pages = {144105},
  issn = {2469-9950, 2469-9969},
  doi = {10.1103/PhysRevB.102.144105},
  urldate = {2022-10-03},
  langid = {english},
}

@article{mackoit-sinkevicieneCarbonDimerDefect2019,
  title = {Carbon Dimer Defect as a Source of the 4.1 {{eV}} Luminescence in Hexagonal Boron           Nitride},
  author = {{Mackoit-Sinkevi{\v c}ien{\.e}}, M. and Maciaszek, M. and {Van de Walle}, C. G. and Alkauskas, A.},
  year = {2019},
  month = nov,
  journal = {Appl. Phys. Lett.},
  volume = {115},
  number = {21},
  pages = {212101},
  publisher = {American Institute of Physics},
  issn = {0003-6951},
  doi = {10.1063/1.5124153},
  urldate = {2021-10-19},
}

@article{tawfikFirstprinciplesInvestigationQuantum2017,
  title = {First-Principles Investigation of Quantum Emission from {{hBN}} Defects},
  author = {Tawfik, Sherif Abdulkader and Ali, Sajid and Fronzi, Marco and Kianinia, Mehran and Tran, Toan Trong and Stampfl, Catherine and Aharonovich, Igor and Toth, Milos and Ford, Michael J.},
  year = {2017},
  month = sep,
  journal = {Nanoscale},
  volume = {9},
  number = {36},
  pages = {13575--13582},
  publisher = {The Royal Society of Chemistry},
  issn = {2040-3372},
  doi = {10.1039/C7NR04270A},
  urldate = {2021-11-30},
}

@article{alkauskasFirstprinciplesTheoryNonradiative2014b,
  title = {First-Principles Theory of Nonradiative Carrier Capture via Multiphonon Emission},
  author = {Alkauskas, Audrius and Yan, Qimin and {Van de Walle}, Chris G.},
  year = {2014},
  month = aug,
  journal = {Phys. Rev. B},
  volume = {90},
  number = {7},
  pages = {075202},
  publisher = {American Physical Society},
  doi = {10.1103/PhysRevB.90.075202},
  urldate = {2024-10-16},
}

@article{alkauskasFirstPrinciplesCalculationsLuminescence2012a,
  title = {First-{{Principles Calculations}} of {{Luminescence Spectrum Line Shapes}} for {{Defects}} in {{Semiconductors}}: {{The Example}} of {{GaN}} and {{ZnO}}},
  shorttitle = {First-{{Principles Calculations}} of {{Luminescence Spectrum Line Shapes}} for {{Defects}} in {{Semiconductors}}},
  author = {Alkauskas, Audrius and Lyons, John L. and Steiauf, Daniel and Van De Walle, Chris G.},
  year = {2012},
  month = dec,
  journal = {Phys. Rev. Lett.},
  volume = {109},
  number = {26},
  pages = {267401},
  issn = {0031-9007, 1079-7114},
  doi = {10.1103/PhysRevLett.109.267401},
  urldate = {2024-10-20},
  copyright = {http://link.aps.org/licenses/aps-default-license},
  langid = {english},
}

@article{frodasonZnVacancydonorImpurity2018,
  title = {Zn Vacancy-Donor Impurity Complexes in {{ZnO}}},
  author = {Frodason, Y. K. and Johansen, K. M. and Bj{\o}rheim, T. S. and Svensson, B. G. and Alkauskas, A.},
  year = {2018},
  month = mar,
  journal = {Phys. Rev. B},
  volume = {97},
  number = {10},
  pages = {104109},
  issn = {2469-9950, 2469-9969},
  doi = {10.1103/PhysRevB.97.104109},
  urldate = {2024-10-20},
  langid = {english},
}

@article{lyonsDeepDonorState2017,
  title = {Deep Donor State of the Copper Acceptor as a Source of Green Luminescence in {{ZnO}}},
  author = {Lyons, J. L. and Alkauskas, A. and Janotti, A. and Van De Walle, C. G.},
  year = {2017},
  month = jul,
  journal = {Appl. Phys. Lett.},
  volume = {111},
  number = {4},
  pages = {042101},
  issn = {0003-6951, 1077-3118},
  doi = {10.1063/1.4995404},
  urldate = {2023-11-30},
  langid = {english},
}

@article{bulancea-lindvallIsotopePurificationInducedReductionSpinRelaxation2023,
  title = {Isotope-{{Purification-Induced Reduction}} of {{Spin-Relaxation}} and {{Spin-Coherence Times}} in {{Semiconductors}}},
  author = {{Bulancea-Lindvall}, Oscar and Eiles, Matthew T. and Son, Nguyen Tien and Abrikosov, Igor A. and Iv{\'a}dy, Viktor},
  year = {2023},
  month = jun,
  journal = {Phys. Rev. Applied},
  volume = {19},
  number = {6},
  pages = {064046},
  issn = {2331-7019},
  doi = {10.1103/PhysRevApplied.19.064046},
  urldate = {2024-12-05},
  langid = {english},
}

@article{fedorovInvestigationIntrinsicDefect2017,
title = {Investigation of intrinsic defect magnetic properties in wurtzite ZnO materials},
journal = {J. Magn. Magn. Mater},
volume = {440},
pages = {5-9},
year = {2017},
issn = {0304-8853},
doi = {https://doi.org/10.1016/j.jmmm.2016.12.130},
url = {https://www.sciencedirect.com/science/article/pii/S0304885316327767},
author = {A.S. Fedorov and M.A. Visotin and A.S. Kholtobina and A.A. Kuzubov and N.S. Mikhaleva and Hua Shu Hsu},

}

@article{leeFirstprinciplesTheoryExtending2022a,
  title = {First-Principles Theory of Extending the Spin Qubit Coherence Time in Hexagonal Boron Nitride},
  author = {Lee, Jaewook and Park, Huijin and Seo, Hosung},
  year = {2022},
  month = sep,
  journal = {npj 2D Mater Appl},
  volume = {6},
  number = {1},
  pages = {1--9},
  publisher = {Nature Publishing Group},
  issn = {2397-7132},
  doi = {10.1038/s41699-022-00336-2},
  urldate = {2024-06-16},
  copyright = {2022 The Author(s)},
  langid = {english},
}

@article{linFirstprinciplesStudyMagnetic2019,
  title = {A First-Principles Study on Magnetic Properties of the Intrinsic Defects in Wurtzite {{ZnO}}},
  author = {Lin, Q. L. and Li, G. P. and Xu, N. N. and Liu, H. and E, D. J. and Wang, C. L.},
  year = {2019},
  month = mar,
  journal = {J. Chem. Phys},
  volume = {150},
  number = {9},
  pages = {094704},
  issn = {0021-9606, 1089-7690},
  doi = {10.1063/1.5063953},
  urldate = {2024-12-05},
  langid = {english},
}

@article{mazeFreeInductionDecay2012,
  title = {Free Induction Decay of Single Spins in Diamond},
  author = {Maze, J. R. and Dr{\'e}au, A. and Waselowski, V. and Duarte, H. and Roch, J.-F. and Jacques, V.},
  year = {2012},
  month = oct,
  journal = {New J. Phys.},
  volume = {14},
  number = {10},
  pages = {103041},
  publisher = {IOP Publishing},
  issn = {1367-2630},
  doi = {10.1088/1367-2630/14/10/103041},
  urldate = {2024-12-05},
  langid = {english},
}

@article{mimsEnvelopeModulationSpinEcho1972,
  title = {Envelope {{Modulation}} in {{Spin-Echo Experiments}}},
  author = {Mims, W. B.},
  year = {1972},
  month = apr,
  journal = {Phys. Rev. B},
  volume = {5},
  number = {7},
  pages = {2409--2419},
  issn = {0556-2805},
  doi = {10.1103/PhysRevB.5.2409},
  urldate = {2024-12-05},
  copyright = {http://link.aps.org/licenses/aps-default-license},
  langid = {english},
}

@article{nagyHighfidelitySpinOptical2019,
  title = {High-Fidelity Spin and Optical Control of Single Silicon-Vacancy Centres in Silicon Carbide},
  author = {Nagy, Roland and Niethammer, Matthias and Widmann, Matthias and Chen, Yu-Chen and Udvarhelyi, P{\'e}ter and Bonato, Cristian and Hassan, Jawad Ul and Karhu, Robin and Ivanov, Ivan G. and Son, Nguyen Tien and Maze, Jeronimo R. and Ohshima, Takeshi and Soykal, {\"O}ney O. and Gali, {\'A}d{\'a}m and Lee, Sang-Yun and Kaiser, Florian and Wrachtrup, J{\"o}rg},
  year = {2019},
  month = apr,
  journal = {Nat Commun},
  volume = {10},
  number = {1},
  pages = {1954},
  issn = {2041-1723},
  doi = {10.1038/s41467-019-09873-9},
  urldate = {2024-12-05},
  langid = {english},
}

@article{parkDecoherenceNitrogenvacancySpin2022,
  title = {Decoherence of Nitrogen-Vacancy Spin Ensembles in a Nitrogen Electron-Nuclear Spin Bath in Diamond},
  author = {Park, Huijin and Lee, Junghyun and Han, Sangwook and Oh, Sangwon and Seo, Hosung},
  year = {2022},
  month = aug,
  journal = {npj Quantum Inf},
  volume = {8},
  number = {1},
  pages = {1--6},
  publisher = {Nature Publishing Group},
  issn = {2056-6387},
  doi = {10.1038/s41534-022-00605-4},
  urldate = {2024-06-16},
  copyright = {2022 The Author(s)},
  langid = {english},
}

@article{roseSpinCoherence14N2017,
  title = {Spin Coherence and {{14N ESEEM}} Effects of Nitrogen-Vacancy Centers in Diamond with {{X-band}} Pulsed {{ESR}}},
  author = {Rose, B. C. and Weis, C. D. and Tyryshkin, A. M. and Schenkel, T. and Lyon, S. A.},
  year = {2017},
  month = feb,
  journal = {Diam. Relat. Mater},
  volume = {72},
  pages = {32--40},
  issn = {0925-9635},
  doi = {10.1016/j.diamond.2016.12.009},
  urldate = {2024-12-05},
}

@article{savchenkoRoleParamagneticDonorlike2020,
  title = {Role of the Paramagnetic Donor-like Defects in the High n-Type Conductivity of the Hydrogenated {{ZnO}} Microparticles},
  author = {Savchenko, Dariya and Vasin, Andrii and Kuz, Oleksandr and Verovsky, Igor and Prokhorov, Andrey and Nazarov, Alexey and Lan{\v c}ok, Jan and Kalabukhova, Ekaterina},
  year = {2020},
  month = oct,
  journal = {Sci Rep},
  volume = {10},
  number = {1},
  pages = {17347},
  issn = {2045-2322},
  doi = {10.1038/s41598-020-74449-3},
  urldate = {2024-12-05},
  langid = {english},
}

@article{seoQuantumDecoherenceDynamics2016,
  title = {Quantum Decoherence Dynamics of Divacancy Spins in Silicon Carbide},
  author = {Seo, Hosung and Falk, Abram L. and Klimov, Paul V. and Miao, Kevin C. and Galli, Giulia and Awschalom, David D.},
  year = {2016},
  month = sep,
  journal = {Nat Commun},
  volume = {7},
  number = {1},
  pages = {12935},
  publisher = {Nature Publishing Group},
  issn = {2041-1723},
  doi = {10.1038/ncomms12935},
  urldate = {2024-06-16},
  copyright = {2016 The Author(s)},
  langid = {english},
}

@article{tuomistoEvidenceZnVacancy2003,
  title = {Evidence of the {{Zn Vacancy Acting}} as the {{Dominant Acceptor}} in n -{{Type ZnO}}},
  author = {Tuomisto, F. and Ranki, V. and Saarinen, K. and Look, D C.},
  year = {2003},
  month = nov,
  journal = {Phys. Rev. Lett.},
  volume = {91},
  number = {20},
  pages = {205502},
  issn = {0031-9007, 1079-7114},
  doi = {10.1103/PhysRevLett.91.205502},
  urldate = {2024-12-05},
  copyright = {http://link.aps.org/licenses/aps-default-license},
  langid = {english},
}

@article{vlasenkoOpticalDetectionElectron2005a,
  title = {Optical Detection of Electron Paramagnetic Resonance in Room-Temperature Electron-Irradiated {{ZnO}}},
  author = {Vlasenko, L. S. and Watkins, G. D.},
  year = {2005},
  month = mar,
  journal = {Phys. Rev. B},
  volume = {71},
  number = {12},
  pages = {125210},
  issn = {1098-0121, 1550-235X},
  doi = {10.1103/PhysRevB.71.125210},
  urldate = {2024-12-05},
  copyright = {http://link.aps.org/licenses/aps-default-license},
  langid = {english},
}

@article{yangElectronSpinDecoherence2014,
  title = {Electron Spin Decoherence in Silicon Carbide Nuclear Spin Bath},
  author = {Yang, Li-Ping and Burk, Christian and Widmann, Matthias and Lee, Sang-Yun and Wrachtrup, J{\"o}rg and Zhao, Nan},
  year = {2014},
  month = dec,
  journal = {Phys. Rev. B},
  volume = {90},
  number = {24},
  pages = {241203},
  issn = {1098-0121, 1550-235X},
  doi = {10.1103/PhysRevB.90.241203},
  urldate = {2024-12-05},
  copyright = {http://link.aps.org/licenses/aps-default-license},
  langid = {english},
}

@article{zhaoDecoherenceDynamicalDecoupling2012,
  title = {Decoherence and Dynamical Decoupling Control of Nitrogen Vacancy Center Electron Spins in Nuclear Spin Baths},
  author = {Zhao, Nan and Ho, Sai-Wah and Liu, Ren-Bao},
  year = {2012},
  month = mar,
  journal = {Phys. Rev. B},
  volume = {85},
  number = {11},
  pages = {115303},
  issn = {1098-0121, 1550-235X},
  doi = {10.1103/PhysRevB.85.115303},
  urldate = {2024-06-16},
  copyright = {http://link.aps.org/licenses/aps-default-license},
  langid = {english},
}

@article{zuoFerromagnetismPureWurtzite2009,
  title = {Ferromagnetism in Pure Wurtzite Zinc Oxide},
  author = {Zuo, Xu and Yoon, Soack-Dae and Yang, Aria and Duan, Wen-Hui and Vittoria, Carmine and Harris, Vincent G.},
  year = {2009},
  month = apr,
  journal = {J. Appl. Phys},
  volume = {105},
  number = {7},
  pages = {07C508},
  issn = {0021-8979, 1089-7550},
  doi = {10.1063/1.3062822},
  urldate = {2024-12-05},
  langid = {english},
}

@article{ivadyFirstPrinciplesCalculation2018,
  title = {First Principles Calculation of Spin-Related Quantities for Point Defect Qubit Research},
  author = {Iv{\'a}dy, Viktor and Abrikosov, Igor A. and Gali, Adam},
  year = {2018},
  month = dec,
  journal = {npj Comput Mater},
  volume = {4},
  number = {1},
  pages = {1--13},
  publisher = {Nature Publishing Group},
  issn = {2057-3960},
  doi = {10.1038/s41524-018-0132-5},
  urldate = {2022-09-23},
  copyright = {2018 The Author(s)},
  langid = {english},

}

@article{liExcitedstateDynamicsOptically2024a,
  title = {Excited-State Dynamics and Optically Detected Magnetic Resonance of Solid-State Spin Defects from First Principles},
  author = {Li, Kejun and Dergachev, Vsevolod D. and Dergachev, Ilya D. and Zhang, Shimin and Varganov, Sergey A. and Ping, Yuan},
  year = {2024},
  month = nov,
  journal = {Phys. Rev. B},
  volume = {110},
  number = {18},
  pages = {184302},
  issn = {2469-9950, 2469-9969},
  doi = {10.1103/PhysRevB.110.184302},
  urldate = {2024-12-30},
  langid = {english}
}

@article{lenefElectronicStructureCenter1996,
  title = {Electronic Structure of the {{N-}} {{{\emph{V}}}} Center in Diamond: {{Theory}}},
  shorttitle = {Electronic Structure of the {{N-}} {{{\emph{V}}}} Center in Diamond},
  author = {Lenef, A. and Rand, S. C.},
  year = {1996},
  month = may,
  journal = {Phys. Rev. B},
  volume = {53},
  number = {20},
  pages = {13441--13455},
  issn = {0163-1829, 1095-3795},
  doi = {10.1103/PhysRevB.53.13441},
  urldate = {2024-10-30},
  copyright = {http://link.aps.org/licenses/aps-default-license},
  langid = {english}
}

@article{dohertyNegativelyChargedNitrogenvacancy2011,
  title = {The Negatively Charged Nitrogen-Vacancy Centre in Diamond: The Electronic Solution},
  shorttitle = {The Negatively Charged Nitrogen-Vacancy Centre in Diamond},
  author = {Doherty, M W and Manson, N B and Delaney, P and Hollenberg, L C L},
  year = {2011},
  month = feb,
  journal = {New J. Phys.},
  volume = {13},
  number = {2},
  pages = {025019},
  issn = {1367-2630},
  doi = {10.1088/1367-2630/13/2/025019},
  urldate = {2024-06-25}
}

@article{bersukerJahnTellerPseudoJahnTeller2017,
  title = {The {{Jahn-Teller}} and Pseudo {{Jahn-Teller}} Effect in Materials Science},
  author = {Bersuker, I. B.},
  year = {2017},
  month = apr,
  journal = {J. Phys.: Conf. Ser.},
  volume = {833},
  number = {1},
  pages = {012001},
  publisher = {IOP Publishing},
  issn = {1742-6596},
  doi = {10.1088/1742-6596/833/1/012001},
  urldate = {2025-01-01},
  langid = {english},
}

@article{razinkovasVibrationalVibronicStructure2021,
  title = {Vibrational and Vibronic Structure of Isolated Point Defects: {{The}} Nitrogen-Vacancy Center in Diamond},
  shorttitle = {Vibrational and Vibronic Structure of Isolated Point Defects},
  author = {Razinkovas, Lukas and Doherty, Marcus W. and Manson, Neil B. and {Van de Walle}, Chris G. and Alkauskas, Audrius},
  year = {2021},
  month = jul,
  journal = {Phys. Rev. B},
  volume = {104},
  number = {4},
  pages = {045303},
  publisher = {American Physical Society},
  doi = {10.1103/PhysRevB.104.045303},
  urldate = {2025-01-01},
}

@article{thieringTheoryOpticalSpinpolarization2018,
  title = {Theory of the Optical Spin-Polarization Loop of the Nitrogen-Vacancy Center in Diamond},
  author = {Thiering, Gerg{\H o} and Gali, Adam},
  year = {2018},
  month = aug,
  journal = {Phys. Rev. B},
  volume = {98},
  number = {8},
  pages = {085207},
  issn = {2469-9950, 2469-9969},
  doi = {10.1103/PhysRevB.98.085207},
  urldate = {2024-12-31},
  langid = {english},
}

@article{streltsovJahnTellerEffectSpinOrbit2020,
  title = {Jahn-{{Teller Effect}} and {{Spin-Orbit Coupling}}: {{Friends}} or {{Foes}}?},
  shorttitle = {Jahn-{{Teller Effect}} and {{Spin-Orbit Coupling}}},
  author = {Streltsov, Sergey V. and Khomskii, Daniel I.},
  year = {2020},
  month = aug,
  journal = {Phys. Rev. X},
  volume = {10},
  number = {3},
  pages = {031043},
  publisher = {American Physical Society},
  doi = {10.1103/PhysRevX.10.031043},
  urldate = {2024-12-31}
}

@book{bersukerJahnTellerEffect2006,
  title = {The {{Jahn-Teller Effect}}},
  author = {Bersuker, Isaac},
  year = {2006},
  publisher = {Cambridge University Press},
  address = {Cambridge},
  doi = {10.1017/CBO9780511524769},
  urldate = {2025-01-01},
}

@article{thieringInitioMagnetoOpticalSpectrum2018,
  title = {{\emph{Ab }}{{{\emph{Initio}}}} {{Magneto-Optical Spectrum}} of {{Group-IV Vacancy Color Centers}} in {{Diamond}}},
  author = {Thiering, Gerg{\H o} and Gali, Adam},
  year = {2018},
  month = jun,
  journal = {Phys. Rev. X},
  volume = {8},
  number = {2},
  pages = {021063},
  issn = {2160-3308},
  doi = {10.1103/PhysRevX.8.021063},
  urldate = {2025-03-20},
  langid = {english},
}

@article{raysonFirstPrinciplesMethod2008b,
  title = {First Principles Method for the Calculation of Zero-Field Splitting Tensors in Periodic Systems},
  author = {Rayson, M. J. and Briddon, P. R.},
  year = {2008},
  month = jan,
  journal = {Phys. Rev. B},
  volume = {77},
  number = {3},
  pages = {035119},
  publisher = {American Physical Society},
  doi = {10.1103/PhysRevB.77.035119},
  urldate = {2024-06-17},

}

@article{neeseCalculationZerofieldSplitting2007,
  title = {Calculation of the Zero-Field Splitting Tensor on the Basis of Hybrid Density Functional and {{Hartree-Fock}} Theory},
  author = {Neese, Frank},
  year = {2007},
  month = oct,
  journal = {J. Chem. Phys},
  volume = {127},
  number = {16},
  pages = {164112},
  issn = {0021-9606},
  doi = {10.1063/1.2772857},
  urldate = {2025-01-21},
}

@article{coffmanSpinHamiltonianParametersSpinOrbit1971,
  title = {Spin-{{Hamiltonian Parameters}} and {{Spin-Orbit Coupling}} for {{V}} 3 + in {{ZnO}}},
  author = {Coffman, R. E. and Himaya, Makram I. and Nyeu, Kathryn},
  year = {1971},
  month = nov,
  journal = {Phys. Rev. B},
  volume = {4},
  number = {9},
  pages = {3250--3252},
  issn = {0556-2805},
  doi = {10.1103/PhysRevB.4.3250},
  urldate = {2025-01-31},
  copyright = {http://link.aps.org/licenses/aps-default-license},
  langid = {english}
}

@article{filipovichElectronParamagneticResonance1970,
  title = {Electron {{Paramagnetic Resonance}} of {{V}} 3 + {{Ions}} in {{Zinc Oxide}}},
  author = {Filipovich, G. and Taylor, A. L. and Coffman, R. E.},
  year = {1970},
  month = mar,
  journal = {Phys. Rev. B},
  volume = {1},
  number = {5},
  pages = {1986--1994},
  issn = {0556-2805},
  doi = {10.1103/PhysRevB.1.1986},
  urldate = {2025-01-31},
  copyright = {http://link.aps.org/licenses/aps-default-license},
  langid = {english}
}

@article{kresseEfficiencyAbinitioTotal1996,
  title = {Efficiency of Ab-Initio Total Energy Calculations for Metals and Semiconductors Using a Plane-Wave Basis Set},
  author = {Kresse, G. and Furthm{\"u}ller, J.},
  year = {1996},
  month = jul,
  journal = {Comput. Mater. Sci},
  volume = {6},
  number = {1},
  pages = {15--50},
  issn = {0927-0256},
  doi = {10.1016/0927-0256(96)00008-0},
  urldate = {2024-05-29},
}

@article{kresseEfficientIterativeSchemes1996,
  title = {Efficient Iterative Schemes for Ab Initio Total-Energy Calculations Using a Plane-Wave Basis Set},
  author = {Kresse, G. and Furthm{\"u}ller, J.},
  year = {1996},
  month = oct,
  journal = {Phys. Rev. B},
  volume = {54},
  number = {16},
  pages = {11169--11186},
  publisher = {American Physical Society},
  doi = {10.1103/PhysRevB.54.11169},
  urldate = {2024-05-29},
}

@article{kresseInitioMoleculardynamicsSimulation1994,
  title = {Ab Initio Molecular-Dynamics Simulation of the Liquid-Metal--Amorphous-Semiconductor Transition in Germanium},
  author = {Kresse, G. and Hafner, J.},
  year = {1994},
  month = may,
  journal = {Phys. Rev. B},
  volume = {49},
  number = {20},
  pages = {14251--14269},
  publisher = {American Physical Society},
  doi = {10.1103/PhysRevB.49.14251},
  urldate = {2024-05-29},
}

@article{heydHybridFunctionalsBased2003,
  title = {Hybrid Functionals Based on a Screened {{Coulomb}} Potential},
  author = {Heyd, Jochen and Scuseria, Gustavo E. and Ernzerhof, Matthias},
  year = {2003},
  month = may,
  journal = {J. Chem. Phys},
  volume = {118},
  number = {18},
  pages = {8207--8215},
  issn = {0021-9606, 1089-7690},
  doi = {10.1063/1.1564060},
  urldate = {2024-12-08},
  langid = {english},
}

@article{giannozziQUANTUMESPRESSOModular2009,
  title = {{{QUANTUM ESPRESSO}}: A Modular and Open-Source Software Project for Quantum Simulations of Materials},
  shorttitle = {{{QUANTUM ESPRESSO}}},
  author = {Giannozzi, Paolo and Baroni, Stefano and Bonini, Nicola and Calandra, Matteo and Car, Roberto and Cavazzoni, Carlo and Ceresoli, Davide and Chiarotti, Guido L and Cococcioni, Matteo and Dabo, Ismaila and Dal Corso, Andrea and {de Gironcoli}, Stefano and Fabris, Stefano and Fratesi, Guido and Gebauer, Ralph and Gerstmann, Uwe and Gougoussis, Christos and Kokalj, Anton and Lazzeri, Michele and {Martin-Samos}, Layla and Marzari, Nicola and Mauri, Francesco and Mazzarello, Riccardo and Paolini, Stefano and Pasquarello, Alfredo and Paulatto, Lorenzo and Sbraccia, Carlo and Scandolo, Sandro and Sclauzero, Gabriele and Seitsonen, Ari P and Smogunov, Alexander and Umari, Paolo and Wentzcovitch, Renata M},
  year = {2009},
  month = sep,
  journal = {J. Phys.: Condens. Matter},
  volume = {21},
  number = {39},
  pages = {395502},
  issn = {0953-8984, 1361-648X},
  doi = {10.1088/0953-8984/21/39/395502},
  urldate = {2022-06-07},
  langid = {english},
}

@article{hamannOptimizedNormconservingVanderbilt2013,
  title = {Optimized Norm-Conserving {{Vanderbilt}} Pseudopotentials},
  author = {Hamann, D. R.},
  year = {2013},
  month = aug,
  journal = {Phys. Rev. B},
  volume = {88},
  number = {8},
  pages = {085117},
  issn = {1098-0121, 1550-235X},
  doi = {10.1103/PhysRevB.88.085117},
  urldate = {2022-06-06},
  langid = {english},
}

@article{schlipfOptimizationAlgorithmGeneration2015,
  title = {Optimization Algorithm for the Generation of {{ONCV}} Pseudopotentials},
  author = {Schlipf, Martin and Gygi, Fran{\c c}ois},
  year = {2015},
  month = nov,
  journal = {Comput. Phys. Commun},
  volume = {196},
  pages = {36--44},
  issn = {0010-4655},
  doi = {10.1016/j.cpc.2015.05.011},
  urldate = {2022-06-06},
  langid = {english},
}

@article{thomasSemiconductorsBasicData1997,
  title = {Semiconductors --- Basic Data, 2nd Ed. {{Edited}} by {{O}}. {{Madelung}}, {{Springer}}, {{Berlin}} 1996, Viii, 317 Pp., Hardcover, {{DM}} 88.00, {{ISBN}} 3-540-60883-4},
  author = {Thomas, Mason},
  year = {1997},
  journal = {Chemical Vapor Deposition},
  volume = {3},
  number = {5},
  pages = {288--289},
  issn = {1521-3862},
  doi = {10.1002/cvde.19970030508},
  urldate = {2024-05-29},
  copyright = {Copyright {\copyright} 1997 Verlag GmbH \& Co. KGaA, Weinheim},
  langid = {english},
}

@article{govoniLargeScaleGW2015,
  title = {Large {{Scale GW Calculations}}},
  author = {Govoni, Marco and Galli, Giulia},
  year = {2015},
  month = jun,
  journal = {J. Chem. Theory Comput.},
  volume = {11},
  number = {6},
  pages = {2680--2696},
  issn = {1549-9618, 1549-9626},
  doi = {10.1021/ct500958p},
  urldate = {2022-12-14},
}

@article{maFirstprinciplesStudiesStrongly2020,
  title = {First-Principles Studies of Strongly Correlated States in Defect Spin Qubits in Diamond},
  author = {Ma, He and Sheng, Nan and Govoni, Marco and Galli, Giulia},
  year = {2020},
  journal = {Phys. Chem. Chem. Phys.},
  volume = {22},
  number = {44},
  pages = {25522--25527},
  issn = {1463-9076, 1463-9084},
  doi = {10.1039/D0CP04585C},
  urldate = {2022-09-11},
  langid = {english}
}

@article{maQuantumSimulationsMaterials2020,
  title = {Quantum Simulations of Materials on Near-Term Quantum Computers},
  author = {Ma, He and Govoni, Marco and Galli, Giulia},
  year = {2020},
  month = jul,
  journal = {npj Comput Mater},
  volume = {6},
  number = {1},
  pages = {1--8},
  publisher = {Nature Publishing Group},
  issn = {2057-3960},
  doi = {10.1038/s41524-020-00353-z},
  urldate = {2022-09-11},
  copyright = {2020 The Author(s)},
  langid = {english},
}

@article{shengGreensFunctionFormulation2022,
  title = {Green's {{Function Formulation}} of {{Quantum Defect Embedding Theory}}},
  author = {Sheng, Nan and Vorwerk, Christian and Govoni, Marco and Galli, Giulia},
  year = {2022},
  month = jun,
  journal = {J. Chem. Theory Comput.},
  volume = {18},
  number = {6},
  pages = {3512--3522},
  issn = {1549-9618, 1549-9626},
  doi = {10.1021/acs.jctc.2c00240},
  urldate = {2022-08-17},
  langid = {english},
}

@article{vorwerkQuantumEmbeddingTheories2022,
  title = {Quantum Embedding Theories to Simulate Condensed Systems on Quantum Computers},
  author = {Vorwerk, Christian and Sheng, Nan and Govoni, Marco and Huang, Benchen and Galli, Giulia},
  year = {2022},
  month = jul,
  journal = {Nat Comput Sci},
  volume = {2},
  number = {7},
  pages = {424--432},
  publisher = {Nature Publishing Group},
  issn = {2662-8457},
  doi = {10.1038/s43588-022-00279-0},
  urldate = {2022-09-11},
  copyright = {2022 Springer Nature America, Inc.},
  langid = {english}
}

@article{neeseORCAProgramSystem2012,
  title = {The {{ORCA}} Program System},
  author = {Neese, Frank},
  year = {2012},
  journal = {Wiley Interdiscip. Rev. Comput. Mol. Sci.},
  volume = {2},
  number = {1},
  pages = {73--78},
  issn = {1759-0884},
  doi = {10.1002/wcms.81},
  urldate = {2024-06-16},
  langid = {english},
}

@article{neeseORCAQuantumChemistry2020,
  title = {The {{ORCA}} Quantum Chemistry Program Package},
  author = {Neese, Frank and Wennmohs, Frank and Becker, Ute and Riplinger, Christoph},
  year = {2020},
  month = jun,
  journal = {J. Chem. Phys},
  volume = {152},
  number = {22},
  pages = {224108},
  issn = {0021-9606},
  doi = {10.1063/5.0004608},
  urldate = {2024-06-16},

}

@article{reiher2006dkh,
  title={{Douglas–Kroll–Hess Theory: a Relativistic Electrons-Only Theory for Chemistry}},
  author={Reiher, Markus},
  journal={Theor. Chem. Acc.},
  volume={116},
  number={},
  pages={241-252},
  year={2006},
  publisher={Springer},
  doi={10.1007/s00214-005-0003-2}
}

@article{rolfes2020all,
  title={All-electron scalar relativistic basis sets for the elements Rb--Xe},
  author={Rolfes, Julian D and Neese, Frank and Pantazis, Dimitrios A},
  journal={J. Comput. Chem.},
  volume={41},
  number={20},
  pages={1842--1849},
  year={2020},
  publisher={Wiley Online Library},
  doi={https://doi.org/10.1002/jcc.26355}
}

@article{pantazis2008all,
  title={All-electron scalar relativistic basis sets for third-row transition metal atoms},
  author={Pantazis, Dimitrios A and Chen, Xian-Yang and Landis, Clark R and Neese, Frank},
  journal={J. Chem. Theory Comput.},
  volume={4},
  number={6},
  pages={908--919},
  year={2008},
  publisher={ACS Publications},
  doi={https://doi.org/10.1021/ct800047t}
}

@article{neese2005soc,
  title={{Efficient and Accurate Approximations to the Molecular Spin-Orbit Coupling Operator and Their Use in Molecular $\emph{g}$-Tensor Calculations}},
  author={Neese, Frank},
  journal={J. Chem. Phys.},
  volume={122},
  number={3},
  pages={034107},
  year={2005},
  publisher={AIP},
  doi={10.1063/1.1829047}
}

@article{weigendBalancedBasisSets2005,
  title = {Balanced Basis Sets of Split Valence, Triple Zeta Valence and Quadruple Zeta Valence Quality for {{H}} to {{Rn}}: {{Design}} and Assessment of Accuracy},
  shorttitle = {Balanced Basis Sets of Split Valence, Triple Zeta Valence and Quadruple Zeta Valence Quality for {{H}} to {{Rn}}},
  author = {Weigend, Florian and Ahlrichs, Reinhart},
  year = {2005},
  journal = {Phys. Chem. Chem. Phys.},
  volume = {7},
  number = {18},
  pages = {3297},
  issn = {1463-9076, 1463-9084},
  doi = {10.1039/b508541a},
  urldate = {2025-01-21},
  langid = {english},
}

@article{badaevaInvestigationPureCo22008,
  title = {Investigation of Pure and {{Co}}{\textsuperscript{2+}} -Doped {{ZnO}} Quantum Dot Electronic Structures Using the Density Functional Theory: Choosing the Right Functional},
  shorttitle = {Investigation of Pure and {{Co}}{\textsuperscript{2+}} -Doped {{ZnO}} Quantum Dot Electronic Structures Using the Density Functional Theory},
  author = {Badaeva, Ekaterina and Feng, Yong and Gamelin, Daniel R and Li, Xiaosong},
  year = {2008},
  month = may,
  journal = {New J. Phys.},
  volume = {10},
  number = {5},
  pages = {055013},
  issn = {1367-2630},
  doi = {10.1088/1367-2630/10/5/055013},
  urldate = {2024-12-31},
  langid = {english},
}

@article{huangSurfacePassivationMethod2005,
  title = {Surface Passivation Method for Semiconductor Nanostructures},
  author = {Huang, Xiangyang and Lindgren, Eric and Chelikowsky, James R.},
  year = {2005},
  month = apr,
  journal = {Phys. Rev. B},
  volume = {71},
  number = {16},
  pages = {165328},
  issn = {1098-0121, 1550-235X},
  doi = {10.1103/PhysRevB.71.165328},
  urldate = {2024-12-31},
  copyright = {http://link.aps.org/licenses/aps-default-license},
  langid = {english}
}

@article{guoSparseMapsSystematicInfrastructure2016,
  title = {{{SparseMaps}}---{{A}} Systematic Infrastructure for Reduced-Scaling Electronic Structure Methods. {{III}}. {{Linear-scaling}} Multireference Domain-Based Pair Natural Orbital {{N-electron}} Valence Perturbation Theory},
  author = {Guo, Yang and Sivalingam, Kantharuban and Valeev, Edward F. and Neese, Frank},
  year = {2016},
  month = mar,
  journal = {The Journal of Chemical Physics},
  volume = {144},
  number = {9},
  pages = {094111},
  issn = {0021-9606, 1089-7690},
  doi = {10.1063/1.4942769},
  urldate = {2025-06-06},
  langid = {english}
}

@article{czelejQuantumBehaviorHydrogenvacancy2018,
  title = {Quantum Behavior of Hydrogen-Vacancy Complexes in Diamond},
  author = {Czelej, Kamil and Zem{\l}a, Marcin Roland and {\'S}piewak, Piotr and Kurzyd{\l}owski, Krzysztof J.},
  year = {2018},
  month = dec,
  journal = {Phys. Rev. B},
  volume = {98},
  number = {23},
  pages = {235111},
  publisher = {American Physical Society},
  doi = {10.1103/PhysRevB.98.235111},
  urldate = {2025-02-14}
}

@article{czelejTransitionMetalRelatedQuantumEmitters2024,
  title = {Transition-{{Metal-Related Quantum Emitters}} in {{Wurtzite AlN}} and {{GaN}}},
  author = {Czelej, Kamil and Lambert, M. Rey and Turiansky, Mark E. and Koshevarnikov, Aleksei and Mu, Sai and {Van de Walle}, Chris G.},
  year = {2024},
  month = oct,
  journal = {ACS Nano},
  volume = {18},
  number = {42},
  pages = {28724--28734},
  publisher = {American Chemical Society},
  issn = {1936-0851},
  doi = {10.1021/acsnano.4c07184},
  urldate = {2025-02-14}
}

@article{galiInitioSupercellCalculations2008,
  title = {Ab Initio Supercell Calculations on Nitrogen-Vacancy Center in Diamond: {{Electronic}} Structure and Hyperfine Tensors},
  shorttitle = {Ab Initio Supercell Calculations on Nitrogen-Vacancy Center in Diamond},
  author = {Gali, Adam and Fyta, Maria and Kaxiras, Efthimios},
  year = {2008},
  month = apr,
  journal = {Phys. Rev. B},
  volume = {77},
  number = {15},
  pages = {155206},
  publisher = {American Physical Society},
  doi = {10.1103/PhysRevB.77.155206},
  urldate = {2025-02-17}
}

@article{shangFirstprinciplesStudyTransition2022,
  title = {First-Principles Study of Transition Metal Dopants as Spin Qubits},
  author = {Shang, Longbing and Chen, Qiaoling and Jing, Weiguo and Ma, Chong-Geng and Duan, Chang-Kui and Du, Jiangfeng},
  year = {2022},
  month = aug,
  journal = {Phys. Rev. Materials},
  volume = {6},
  number = {8},
  pages = {086201},
  issn = {2475-9953},
  doi = {10.1103/PhysRevMaterials.6.086201},
  urldate = {2024-10-01},
  langid = {english},
}

@article{ivanovElectronicExcitationsCharged2023,
  title = {Electronic Excitations of the Charged Nitrogen-Vacancy Center in Diamond Obtained Using Time-Independent Variational Density Functional Calculations},
  author = {Ivanov, Aleksei V. and Schmerwitz, Yorick L. A. and Levi, Gianluca and J{\'o}nsson, Hannes},
  year = {2023},
  month = jul,
  journal = {SciPost Phys.},
  volume = {15},
  number = {1},
  pages = {009},
  issn = {2542-4653},
  doi = {10.21468/SciPostPhys.15.1.009},
  urldate = {2025-02-14},
  langid = {english}
}

@article{liCarbonTrimerEV2022,
  title = {Carbon Trimer as a 2 {{eV}} Single-Photon Emitter Candidate in Hexagonal Boron Nitride: {{A}} First-Principles Study},
  shorttitle = {Carbon Trimer as a 2 {{eV}} Single-Photon Emitter Candidate in Hexagonal Boron Nitride},
  author = {Li, Kejun and Smart, Tyler J. and Ping, Yuan},
  year = {2022},
  month = apr,
  journal = {Phys. Rev. Mater.},
  volume = {6},
  number = {4},
  pages = {L042201},
  publisher = {American Physical Society},
  doi = {10.1103/PhysRevMaterials.6.L042201},
  urldate = {2022-08-08},
}

@article{sangalliManybodyPerturbationTheory2019,
  title = {Many-Body Perturbation Theory Calculations Using the Yambo Code},
  author = {Sangalli, D. and Ferretti, A. and Miranda, H. and Attaccalite, C. and Marri, I. and Cannuccia, E. and Melo, P. and Marsili, M. and Paleari, F. and Marrazzo, A. and Prandini, G. and Bonf{\`a}, P. and Atambo, M. O. and Affinito, F. and Palummo, M. and {Molina-S{\'a}nchez}, A. and Hogan, C. and Gr{\"u}ning, M. and Varsano, D. and Marini, A.},
  year = {2019},
  month = may,
  journal = {J. Phys.: Condens. Matter},
  volume = {31},
  number = {32},
  pages = {325902},
  publisher = {IOP Publishing},
  issn = {0953-8984},
  doi = {10.1088/1361-648X/ab15d0},
  urldate = {2020-12-11},
}

@article{yangQuantumManybodyTheory2008a,
  title = {Quantum Many-Body Theory of Qubit Decoherence in a Finite-Size Spin Bath},
  author = {Yang, Wen and Liu, Ren-Bao},
  year = {2008},
  month = aug,
  journal = {Phys. Rev. B},
  volume = {78},
  number = {8},
  pages = {085315},
  issn = {1098-0121, 1550-235X},
  doi = {10.1103/PhysRevB.78.085315},
  urldate = {2024-12-08},
  copyright = {http://link.aps.org/licenses/aps-default-license},
  langid = {english},
}

@article{AlbertssonZnOExperiment,
author = {Albertsson, J. and Abrahams, S. C. and Kvick, \AA.},
title = {Atomic displacement, anharmonic thermal vibration, expansivity and pyroelectric coefficient thermal dependences in ZnO},
journal = {Acta Crystallographica Section B},
volume = {45},
number = {1},
pages = {34-40},
doi = {https://doi.org/10.1107/S0108768188010109},
year = {1989}
}

@article{ObaDefectEnergetics,
  title = {Defect energetics in ZnO: A hybrid Hartree-Fock density functional study},
  author = {Oba, Fumiyasu and Togo, Atsushi and Tanaka, Isao and Paier, Joachim and Kresse, Georg},
  journal = {Phys. Rev. B},
  volume = {77},
  issue = {24},
  pages = {245202},
  numpages = {6},
  year = {2008},
  month = {Jun},
  publisher = {American Physical Society},
  doi = {10.1103/PhysRevB.77.245202},
  url = {https://link.aps.org/doi/10.1103/PhysRevB.77.245202}
}

@article{ReynoldsZnOExperimentBandgap,
  title = {Valence-band ordering in ZnO},
  author = {Reynolds, D. C. and Look, D. C. and Jogai, B. and Litton, C. W. and Cantwell, G. and Harsch, W. C.},
  journal = {Phys. Rev. B},
  volume = {60},
  issue = {4},
  pages = {2340--2344},
  numpages = {0},
  year = {1999},
  month = {Jul},
  publisher = {American Physical Society},
  doi = {10.1103/PhysRevB.60.2340},
  url = {https://link.aps.org/doi/10.1103/PhysRevB.60.2340}
}

\end{document}